\documentclass[a4paper,14pt, fleqn,usenatbib]{mnras}

\usepackage[T1]{fontenc}
\usepackage{ae,aecompl}

\usepackage{amsmath}	
\usepackage{amssymb}	
\usepackage{lscape}
\usepackage{graphicx, color}

\title[Low-mass star formation in the first galaxies]
{Condition for low-mass star formation in shock-compressed metal-poor clouds}

\author[Nakauchi et al.]{
Daisuke Nakauchi$^{1}$\thanks{E-mail: nakauchi@astr.tohoku.ac.jp},
Kazuyuki Omukai$^{1}$, and Raffaella Schneider$^{2, 3}$
\\
$^{1}$Astronomical Institute, Tohoku University, Aoba, Sendai 980-8578, Japan \\
$^{2}$Dipartimento di Fisica, Universit\'a di Roma `La Sapienza', P.le Aldo Moro 2, I-00185 Roma, Italy\\
$^{3}$INAF/Osservatorio Astronomico di Roma, Via Frascati 33, 00078, Monte Porzio Catone, Roma}
\date{Accepted XXX. Received YYY; in original form ZZZ}

\pubyear{2018}

\begin{document}
\label{firstpage}
\pagerange{\pageref{firstpage}--\pageref{lastpage}}
\maketitle

\begin{abstract}

Shocks may have been prevalent in the early Universe, associated with virialization and supernova explosions, etc.
Here, we study thermal evolution and fragmentation of shock-compressed clouds,
by using a one-zone model with detailed thermal and chemical processes.
We explore a large range of initial density~($1\mbox{-}10^5\ {\rm cm}^{-3}$), metallicity~($0\mbox{-}10^{-2}\ {\rm Z}_\odot$), 
UV strength~($0\mbox{-}500$ times Galactic value), and cosmic microwave background temperature~(10 and 30 K).
Shock-compressed clouds contract isobarically via atomic and molecular line cooling,
until self-gravitating clumps are formed by fragmentation.
If the metals are only in the gas-phase, the clump mass is higher than $\sim 3\ {\rm M}_\odot$ in any conditions we studied.
Although in some cases with a metallicity higher than $\sim 10^{-3}\ {\rm Z}_\odot$, re-fragmentation of a clump is caused by metal-line cooling, 
this fragment mass is higher than $\sim 30\ {\rm M}_\odot$.
On the other hand, if about half the mass of metals is condensed in dust grains, as in the Galactic interstellar medium,
dust cooling triggers re-fragmentation of a clump into sub-solar mass pieces,
for metallicities higher than $\sim 10^{-5}\ {\rm Z}_\odot$.
Therefore, the presence of dust is essential in low-mass~($\lesssim {\rm M}_\odot$) star formation
from a shock-compressed cloud.
\end{abstract}



\section{Introduction}\label{sec:intro}
Theories predict that the first stars formed from primordial pristine gas, the Population III~(Pop III) stars, 
were typically massive with $10\mbox{-}1000\ {\rm M}_\odot$~\citep[e.g.,][]{Bromm2002, Abel2002, Yoshida2008, Hosokawa2012, Hirano2014, Susa2014},
while Pop I stars in the solar neighborhood have a peak in their initial mass function (IMF) around 
$0.1\mbox{-}1\ {\rm M}_\odot$~\citep[e.g.,][]{Kroupa2002,Bastian2010}.
Although the IMF of metal-poor Pop II stars is still uncertain, some of them are surely low-mass objects as 
discovered in the Galactic halo, globular clusters, and nearby dwarf galaxies~\citep[e.g,][]{Beers2005,Frebel2015}.
This indicates a transition from high-mass to low-mass star-formation mode, so-called the Pop III/II transition, over cosmic history.
Such a transition might have been caused by the increasing contribution of turbulence, magnetic fields and metals in star formation.
Metals are known to affect the thermal properties of a star-forming gas by elevating its cooling efficiency and thereby reducing the cloud fragmentation mass scale~\citep[e.g.,][]{Omukai2000}.

Metals can exist in two forms, i.e., in the gas-phase or in dust grains.
For metallicities higher than $Z/{\rm Z}_{\odot} = 10^{-4}\mbox{-}10^{-3}$, fine-structure line cooling by gas-phase metals enables the gas to reach lower temperatures~(at a density of $\sim 10^{4}\ {\rm cm}^{-3}$) than in the primordial case,
lowering the fragmentation mass~\citep{Bromm2001, Bromm_Loeb2003b, Santoro2006, Jappsen2007, Glover2014}.
This cooling is, however, efficient only at rather low densities~($\sim 10^4\ {\rm cm}^{-3}$) and leads to fragments 
with mass scales no smaller than a few $10\ {\rm M}_\odot$.
If some fraction of the metals are in dust grains, as in the Galactic interstellar medium~(ISM), dust cooling triggers fragmentation 
at high enough densities~($\gtrsim 10^{10}\ {\rm cm}^{-3}$) to produce sub-solar mass fragments 
for metallicities exceeding $Z/{\rm Z}_{\odot}=10^{-6}\mbox{-}10^{-5}$~\citep{Schneider2002, Schneider2003, Omukai2005, Schneider2006, Omukai2010, Schneider2010,Schneider2012, Schneider2012b}.
Three-dimensional hydrodynamic calculations also demonstrate the formation of multiple low-mass stars through dust-induced fragmentation for $Z/{\rm Z}_{\odot} \sim 10^{-5}$~\citep{Tsuribe2006, Clark2008, Dopcke2013, Chiaki2016, Bovino2016}.
Therefore, dust is considered to be indispensable to form low-mass ($\lesssim {\rm M}_{\sun}$) stars.

Shocks are presumably prevalent also in the early galaxies, associated with virialization, 
galaxy mergers, expansion of HII regions, outflows, SN explosions etc.
For example, numerical simulations demonstrate that inflows of cold gas penetrate deep in a forming galaxy 
and create shocks in a dense region~\citep{Wise2007,Wise2008, Greif2008, Safranek-Shrader2014a, Safranek-Shrader2016}.
The fragmentation mass-scale in such a shock-compressed layer of low metallicity gas 
was studied by \citet{Safranek-Shrader2010}, using a semi-analytical method.
They calculated the post-shock thermal evolution starting from the initial density and temperature of
$n_{\rm H, 1} = 4 \times 10^3\ {\rm cm}^{-3}$ and $T_1 = 1.1 \times 10^4\ {\rm K}$, suggested by the result of a numerical simulation
by \citet{Greif2008}.
They found that the temperature reaches the CMB floor via metal-line cooling and  
moderately low-mass fragments~($\sim 3\ {\rm M}_\odot$) are formed even without dust, 
if the metallicity is higher than $Z/{\rm Z}_{\odot} \simeq 10^{-2.5}$.

If the star-formation efficiency from the parent dense cores to the stars is about 30\%, 
as suggested by observations in the solar neighborhood, or smaller,
this implies the possibility to form solar or sub-solar mass stars even in the absence of dust cooling.
Their calculation, however, lacked some important thermal processes, like line cooling by metallic molecules~(e.g., H$_2$O, OH, and CO), cooling by dust thermal emission, and heating associated with H$_2$ formation.
The evolution of a self-gravitating clump after fragmentation was not followed in their work either.
In addition, a restricted set of physical conditions~(initial density, UV strength, and CMB temperature) was investigated, 
limiting our ability to understand the prevalence of shock-induced low-mass star formation. 

Here, by modeling the thermal and chemical processes in more detail, we study thermal evolution and fragmentation of 
shock-compressed clouds, for a large range of initial density, metallicity, UV strength, and CMB temperature.
Our aim is to pin down the condition for low-mass star formation.
We find that the mass of a fragment formed by shock-compression does not become smaller than $\simeq 3\ {\rm M}_\odot$
in any environmental conditions we explored.
If the metallicity is higher than $\sim 10^{-5}\ {\rm Z}_\odot$ and a similar fraction of metals as in the Milky Way 
condenses into dust grains, dust cooling enables re-fragmentation into
sub-solar mass dense cores, irrespective of the initial density and UV strength.
Low-mass star formation is possible only in shock-compressed clouds which contain dust grains and have metallicity higher than $\sim 10^{-5}\ {\rm Z}_\odot$.

The rest of the paper is organized as follows.  
In Section \ref{sec:method}, we describe the method to calculate the thermal evolution 
of a shock-compressed cloud leading to fragmentation,  
as well as the subsequent gravitational contraction of the fragments.
Results for thermal evolution are presented in Section \ref{sec:thermal_evolution}, where we also discuss 
their dependences on the UV strength and CMB temperature.
We focus on two specific sets of initial conditions that are expected in early galaxies, 
i.e., cold accretion flows and SN explosions.
Fragmentation mass scales are summarized in Section \ref{sec:fragment_mass}.
We also discuss the condition for low-mass star formation, by seeing the fragment mass formed
under various conditions.
Section \ref{sec:discussion} is devoted to summary and discussion. 

\section{Method}
\label{sec:method}

\subsection{Post-shock Thermal Evolution}
\label{subsec:shock}

We calculate the thermal evolution of a shock-compressed layer under the assumption of a plane-parallel and steady flow~\citep{Shapiro1987,Yamada1998,Inayoshi2012,Nakauchi2014}.
The physical quantities are functions only of the distance $x$ from the shock.  
This treatment corresponds to calculating a spatially averaged profile on the $x=$ const. surface
although the post-shock medium can be inhomogeneous by thermal instabilities etc. 

Density $\rho$, velocity $v$, and pressure $P$ are followed by using the mass and momentum conservations:
\begin{equation}
\rho_1 v_1 = \rho v,
\label{eq:EoC}
\end{equation}
and
\begin{equation}
\rho_1 v_1^2 + P_1 = \rho v^2 + P,
\label{eq:EoM}
\end{equation}
where the subscript `1' indicates the quantities immediately behind the shock front.
The internal energy $e$ per unit mass is obtained by solving the energy equation:
\begin{equation}
v \frac{de}{dx} = - P v \frac{d}{dx}\left(\frac{1}{\rho}\right) - \Lambda_{\rm net},   
\label{eq:EoE_steady}
\end{equation}
where $\Lambda_{\rm net}$~(erg g$^{-1}$ s$^{-1}$) is the net cooling rate, 
which is supplemented by the relations $P = \rho k_{\rm B} T/ \mu m_{\rm H}$ and $e = P/\rho/(\gamma-1)$. 
Here, the symbols have their usual meanings.

For the cooling rate, $\Lambda_{\rm net}$, we consider the following cooling/heating processes: 
radiative cooling of atomic and molecular lines~($\Lambda_{\rm line}$), 
and of continuum from gas and dust~($\Lambda_{\rm cont}$ and $\Lambda_{\rm grain}$), 
heating/cooling associated with chemical reactions~($\Lambda_{\rm chem}$),
and photoelectric heating by dust grains~($\Gamma_{\rm pe}$):
\begin{equation}
\Lambda_{\rm net} = \Lambda_{\rm line} + \Lambda_{\rm cont} + \Lambda_{\rm grain}+ \Lambda_{\rm chem} 
- \Gamma_{\rm pe}.
\label{eq:Lambda_net}
\end{equation}
Line cooling~($\Lambda_{\rm line}$) is contributed by H Ly$\alpha$, the ro-vibrational lines of 
H$_2$, HD, H$_2$O, OH, and CO, and the fine-structure lines of [OI], [CII], [CI], [SiII], and [FeII].
The cooling and heating rates are calculated following \cite{Omukai2010} for H$_2$O, OH, and CO cooling, 
\cite{Schneider2012} for [SiII] and [FeII] cooling, \cite{Omukai2008} for photoelectric heating $\Gamma_{\rm pe}$,
and \cite{Omukai2005} for the remaining processes except for the [OI] and [CI] cooling rates, which are updated by using the collisional excitation rates for hydrogen impact computed by \cite{Abrahamsson2007}.
Reduction of cooling/heating rates due to photon trapping is taken into account by considering the frequency shift due to velocity gradient:
the line optical depth is evaluated from the column density across the layers whose velocity difference is less than the thermal velocity.

Following \cite{Omukai2012}, we calculate the H, He, D, C and O chemical network among the 50 species: H, H$_2$, e$^-$, H$^+$, H$_2^+$, H$_3^+$, H$^-$, He, He$^+$, He$^{2+}$, HeH$^+$, D, HD, D$^+$, HD$^+$, D$^-$, C, C$_2$, CH, CH$_2$, CH$_3$, CH$_4$, C$^+$, C$_2^+$, CH$^+$, CH$_2^+$, CH$_3^+$, CH$_4^+$, CH$_5^+$, O, O$_2$, OH, CO, H$_2$O, HCO, O$_2$H, CO$_2$, H$_2$CO, H$_2$O$_2$, O$^+$, O$_2^+$, OH$^+$, CO$^+$, H$_2$O$^+$, HCO$^+$, O$_2$H$^+$, H$_3$O$^+$, H$_2$CO$^+$, HCO$_2^+$, and H$_3$CO$^+$.
The Si and Fe chemistries are not solved, and we assume these elements to be always in the form of SiII and FeII.

All the photo-reaction rates are scaled by a single parameter, the so-called Habing parameter $G_0$, 
defined as the energy density between 6-13.6 eV, 
normalized by the Galactic field $5.29 \times 10^{-14}\ {\rm erg}\ {\rm cm}^{-3}$~\citep{Habing1968}.
Here, the UV spectrum is implicitly assumed to be the same as the local interstellar field.
The Habing parameter $G_0$ is related to another conventional UV parameter, $J_{21}$, 
the intensity at the Lyman limit normalized by $10^{-21}\ {\rm erg}\ {\rm s}^{-1}\ {\rm cm}^{-2}\ {\rm Hz}^{-1}\ {\rm str}^{-1}$ as
\begin{equation}
J_{21} = 20.9\ G_0,
\label{eq:J21_G0}
\end{equation}
for the spectrum of the local interstellar field~\citep{Mathis1983}.
This relation depends on the spectral shape.
For example, $J_{21} = 0.667\ G_0$ ($102\ G_0$) for a black-body spectrum with $10^4$ K~($10^5$ K).
Among dozens of photo-reactions included in our calculation, the two most relevant to thermal evolution 
are H$_2$ photo-dissociation and C photoionization, both of which are controlled by the radiation in the frequency range $\simeq$ 11-12 eV.
If the value of $J_{21}$ is same, the intensity at $\simeq 11$ eV for the black body spectrum with $10^4$ K~($10^5$ K)
is larger by a factor of 3.7 (0.27) compared to the \cite{Mathis1983} spectrum.
Therefore, our results calculated with a certain UV strength of $J_{21}$ reproduce those calculated with $0.27\ J_{21}$~($3.7\ J_{21}$)
for the black body spectrum with $10^4$ K~($10^5$ K).

\subsection{Fragmentation Criterion of a Shock-Compressed Layer}\label{sec:fragcond}

Here, we consider the condition for fragmentation of the shock-compressed layers, which produces self-gravitating clumps.
\cite{Elmegreen1978} performed a linear perturbation analysis for a plane-parallel layer in hydrostatic equilibrium 
and bounded from both sides by external pressure.
They find that the most unstable perturbation mode of approximately the layer width grows roughly in the free-fall time scale, 
$t_{\rm ff} \equiv \sqrt{{3 \pi}/{32 G \rho}}$, resulting in fragmentation of the layer.
In our model, we also need to take the contraction of the layer into account.
The steady post-shock flow cools almost isobarically and contracts
in the dynamical time scale which is comparable to the cooling time:
from the isobaricity and Eq. \ref{eq:EoE_steady}, 
$t_{\rm dyn} \equiv \rho/\dot{\rho} = \gamma (e / \Lambda_{\rm net}) = \gamma t_{\rm cool} \sim t_{\rm cool}$.
Thus if $t_{\rm cool} < t_{\rm ff}$, the layer contracts further before the perturbations grow and no fragmentation occurs.
Conversely, if $t_{\rm cool} > t_{\rm ff}$, the perturbations have enough time to grow and the shocked layer fragments before further contraction.
These fragments are, however, not self-gravitating
unless the free-fall time $t_{\rm ff}$ is shorter than the sound crossing time $t_{\rm sound}$ across the layer.
We thus adopt the following inequalities in time-scales as the fragmentation condition:
\begin{equation}
t_{\rm cool} > t_{\rm ff}\ \text{and}\ t_{\rm sound} > t_{\rm ff}.
\label{eq:frag_cond}
\end{equation}
Here, the sound crossing time is evaluated by $t_{\rm sound} = H_\rho / c_{\rm s}$,
using the density scale height of the coolest layer $H_\rho = \rho / (d \rho/dr)$ and its sound speed $c_{\rm s} = (\gamma P / \rho)^{1/2}$.
Note that \cite{Safranek-Shrader2010} adopted only the second condition~($t_{\rm sound} > t_{\rm ff}$) as the criterion of fragmentation,
which indicates that the layer becomes self-gravitating.
The first condition~($t_{\rm cool} > t_{\rm ff}$) assures that perturbations have enough time to grow.

\subsection{Gravitational Contraction of Clumps after Fragmentation}

After fragmentation, the clumps continue to contract by self-gravity. 
For spherically symmetric clumps without turbulence or magnetic fields, 
hydrodynamical studies tell us that the actual contraction follows the Larson-Penston type self-similar solution, 
where the central density increases in the local free-fall time scale~\citep{Larson1969, Penston1969}.
We follow its central evolution using a one-zone model: 
\begin{equation}
\frac{d \rho}{dt} = \frac{\rho}{t_{\rm ff}},
\label{eq:free_fall}
\end{equation}
and
\begin{equation}
\frac{de}{dt} = - P \frac{d}{dt}\left(\frac{1}{\rho}\right) - \Lambda_{\rm net},   
\label{eq:EoE}
\end{equation}
coupled with the same cooling rate as in Eq. \eqref{eq:Lambda_net}.
Since the core size is comparable to the local Jeans length, defined by $\lambda_{\rm J} \equiv (\pi k_{\rm B} T / G \mu m_{\rm H} \rho)^{1/2}$, 
photon trapping and shielding are evaluated by using the column density across the core: $N_{\rm H, core} = n_{\rm H} \lambda_{\rm J}$. 

\subsection{Fragmentation Mass Scale}\label{subsec:method_frag_mass}

The clump mass at its formation by fragmentation can be evaluated from the Jeans mass, $M_{\rm J}(\rho, T)$,
by using the density $\rho_{\rm frag}$ and temperature $T_{\rm frag}$ at that epoch:
\begin{equation}
M_{\rm J}(\rho_{\rm frag}, T_{\rm frag}) \simeq \rho_{\rm frag} \lambda_{\rm J}^3(\rho_{\rm frag}, T_{\rm frag}). 
\label{eq:jeans_mass}
\end{equation}

For the clump to continue its gravitational contraction, its mass should always exceed 
the instantaneous Jeans mass $M_{\rm J}(\rho, T)$.
If the temperature increases so rapidly that the effective ratio of specific heat $\gamma_{\rm eff}~(\equiv d \ln P / d \ln \rho$)
exceeds 4/3, the instantaneous Jeans mass increases with contraction and can eventually exceed the clump mass.
In such cases, the clump would contract in a quasi-static manner by accreting surrounding material,
keeping its mass roughly at the instantaneous Jeans mass.
Therefore, the final clump mass is given by the maximum value of the instantaneous Jeans mass $M_{\rm J}(\rho, T)$ 
reached during gravitational contraction: $M_{\rm clump} = {\rm max} \left[M_{\rm J}(\rho, T)\right]$.

Re-fragmentation of a clump can occur during gravitational contraction, if the clump experiences a rapid cooling phase. 
Following \cite{Schneider2010}, we assume that fragmentation occurs if the temperature drops suddenly with $\gamma_{\rm eff} < 0.8$.
The fragment mass is set by the Jeans mass when the cooling phase is almost over with $\gamma_{\rm eff} \geq 0.97$. 
Such fragmentation can be driven either by line or dust cooling~\citep{Schneider2002, Omukai2005, Tsuribe2006, Tsuribe2008}.
We call this fragment mass scale as $M_{\rm re-frag}$.

\subsection{Initial Settings}

We study the evolution of a shock-compressed cloud, focusing mainly on two situations.
The first is a cold accretion~(CA) flow penetrating into a protogalaxy with a speed of $\sim 20\ {\rm km}\ {\rm s}^{-1}$.
Shocks associated with it heat the medium in the dense region of a protogalaxy,
whose typical density is $\sim 10^3\ {\rm cm}^{-3}$~\citep{Wise2007,Wise2008, Greif2008, Safranek-Shrader2010}.
Therefore, in the CA case, the initial density and temperature are set at $n_{\rm H, 1} = 4 \times 10^3\ {\rm cm}^{-3}$
and $T_1 = 1.2 \times 10^4\ {\rm K}$, respectively.
The second is a SN shock propagating into an HII region with $\sim 0.1\mbox{-}1\ {\rm cm}^{-3}$,
which surrounds the progenitor massive star~\citep{Kitayama2005, Nagakura2009, Chiaki2013}.
In the early snow-plough phase of a SN remnant~(SNR), the temperature of a layer immediately behind the shock front
becomes higher than $\sim 10^5\ {\rm K}$ due to the large shock velocities of $\sim 50\mbox{-100}\ {\rm km}\ {\rm s}^{-1}$.
It soon, however, decreases down to $\sim 10^4\ {\rm K}$ via H and He atomic cooling.
In the SN case, we start the calculation from a post-shock layer that has already compressed to 
$n_{\rm H, 1} = 4 \ {\rm cm}^{-3}$ and $T_1 = 1.2 \times 10^4\ {\rm K}$.
Considering the possibility that a SN shock sweeps up an ISM denser than $\sim 0.1\mbox{-}1\ {\rm cm}^{-3}$,
we explore a larger range of initial densities with $1\ {\rm cm}^{-3} < n_{\rm H, 1} < 10^5\ {\rm cm}^{-3}$.
In the snow plough phase, the width of the post shock layer is much smaller than the radius of
a shock front and the curvature of the shock is negligible, as shown by the hydrodynamical
calculation of a SNR evolution~\citep{Nagakura2009}. Therefore, the plane-parallel shock model
is also applicable to the SN case.

We study metal-poor clouds in the metallicity range of $0 < Z/{\rm Z}_{\odot} \leq 10^{-2}$,
and set the CMB temperature at $T_{\rm CMB} = 30 \ {\rm K}$.
The case of a lower CMB temperature, $T_{\rm CMB} = 10 \ {\rm K}$, is also studied for comparison.
The UV strength in early galaxies is highly uncertain.
If stars are formed much more actively in the early galaxies than in the solar neighborhood, 
the UV strength can be much higher than the local interstellar value, $J_{21} \simeq 20~(G_0 \simeq 1)$~\citep{Mathis1983}.
Therefore, we explore a broad range of UV strengths: $0 \leq J_{21} \leq 10^4$.

The initial fractions of H$^+$, H$_2$, and HD are set as $y({\rm H}^+)_{\rm ini} = 10^{-4}$, $y({\rm H}_2)_{\rm ini} = 10^{-6}$, and $y({\rm HD})_{\rm ini} = 10^{-9}$, respectively~\citep[e.g.,][]{Galli2013}.
The other H and D nuclei are in the neutral atoms.
The elemental abundances of D and He are $y_{\rm D} = 2.5 \times 10^{-5}$ and $y_{\rm He} = 8.3 \times 10^{-2}$, respectively.
The amount of metals depleted into dust grains affect the metal abundances in the gas-phase.
In principle, the relative abundances of metals and dust should depend on the chemical enrichment history 
of the galaxy~\citep{Schneider2006, Schneider2012b, Marassi2015, deBennassuti2014, deBennassuti2017, Ginolfi2018}.
However, here we ignore this complication and simply assume that, when no dust is present, 
the gas-phase metal composition is proportional to the solar abundance pattern~\citep{Anders1989}.
On the other hand, when dust is present, we take the gas-phase metal composition and dust-to-gas ratio 
to follow the values of local ISM, i.e, 72, 46, 90, and 100\% of C, O, Si, and Fe are depleted into dust grains as modeled by \cite{Pollack1994}.
Since the initial temperature is $\simeq 10^4\ {\rm K}$, all the He, C and O are supposed to be in HeI, CII and OI.

The initial abundances of H$^+$, H$_2$ and HD, adopted in our calculation, are the values expected in the high-redshift intergalactic medium,
and may not be appropriate for a CA and a SN shock.
Without an external ionization source, the recombination proceeds in a free-fall time to $y({\rm H}^+) \sim (10^{-4}\mbox{-}10^{-3}) \times n_{\rm H}^{-1/2}$
in a warm neutral medium of several 1000 K, even with a higher ionization fraction initially.
Although this value is more or less similar to our adopted abundance for the relatively diffuse case with $n_{\rm H} \sim 10\ {\rm cm}^{-3}$,
the initial ionization fraction can differ in the cases with a largely different initial density.
However, we believe our results to be robust due to the following considerations.
Even with a smaller initial ionization degree, the high initial temperature of $\gtrsim$ 10000 K leads to a rapid increase of the ionization fraction to $y({\rm H}^+) \sim 4 \times 10^{-4}$ via collisional ionization, erasing the memory of the initial low ionization degree.
A higher value of ionization degree also has only a minor effect.
The ionization fraction decreases by recombination to a similar level as in the low $y({\rm H}^+)_{\rm ini}$ cases
during the isobaric contraction stage, whose track is solely determined by the shock velocity.
Thus, different ionization degrees do not largely alter our thermal tracks.
In addition, H$_2$ and HD initially present are collisionally dissociated due to the high initial temperature~($\sim$ 10000 K)
and their abundances decrease rapidly to $y({\rm H}_2) \sim 10^{-7}$ and $y({\rm HD}) \sim 10^{-11}$, respectively. 
By the time the gas cools to $\sim$ 1000 K, H$_2$ and HD are formed again with their abundances reaching $y({\rm H}_2) \sim 10^{-3}$ and $y({\rm HD}) \sim 10^{-7}$, respectively.
Therefore, $y({\rm H}_2)$ and $y({\rm HD})$ values in the subsequent stages do not depend on their initial values.

\section{Thermal Evolution}\label{sec:thermal_evolution}
In this section, we describe the thermal evolution of a shock-compressed cloud.
In Sections \ref{subsec:gasflow} and \ref{subsec:snremnant}, we present the results 
for CA and SN shocks, respectively, in the case of no UV irradiation~($J_{21} = 0$). 
In Sections \ref{subsec:UVfield} and \ref{subsec:CMBfield}, we see how the results are altered
by the presence of UV irradiation and by a lower CMB temperature, respectively.

\subsection{Cloud Compressed by a Cold Accretion Shock}~\label{subsec:gasflow}

\begin{figure*}
\centering
\begin{tabular}{c}
{\includegraphics[scale=1.15]{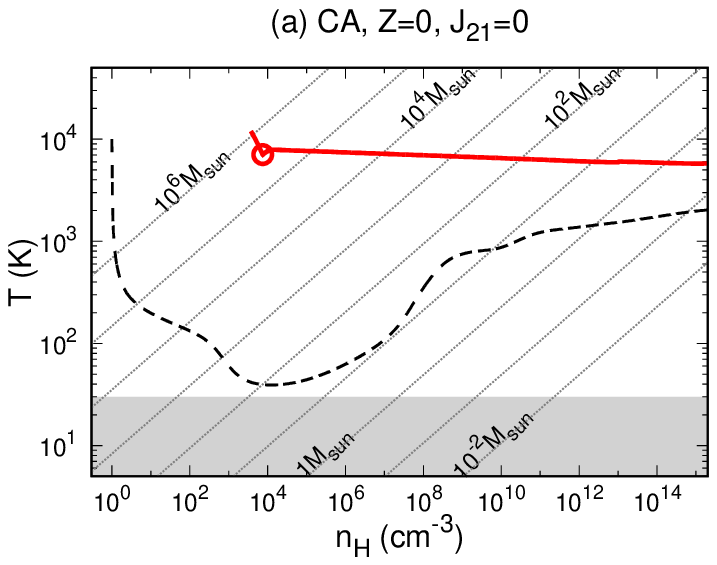}} 
{\includegraphics[scale=1.15]{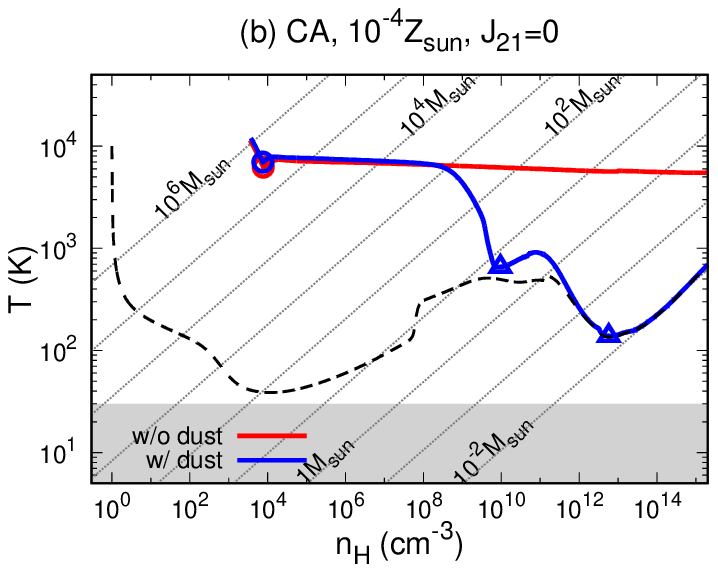}} \\
{\includegraphics[scale=1.15]{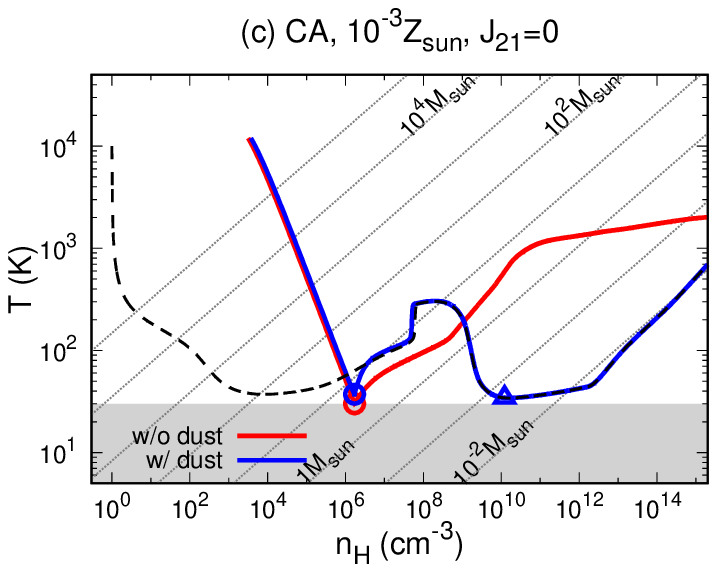}} 
{\includegraphics[scale=1.15]{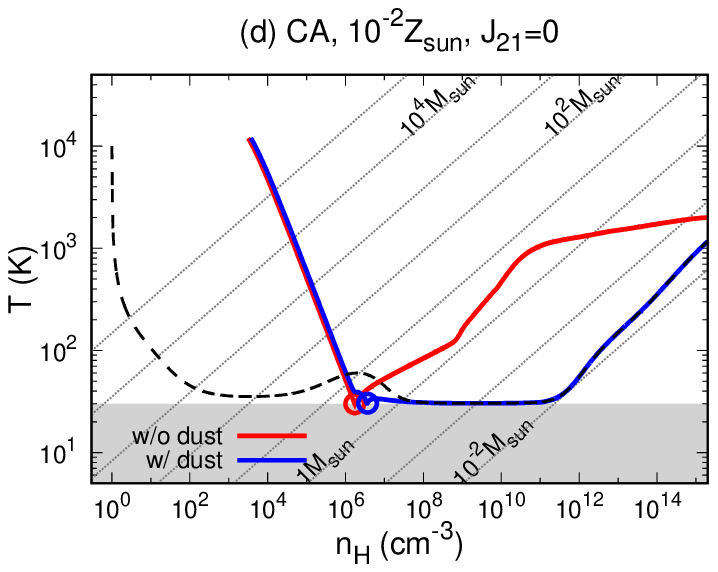}} \\
\end{tabular}
\caption{Thermal evolution of a cloud compressed by a CA shock.
Panels (a), (b), (c), and (d) correspond to the cases with initial metallicities of $Z/{\rm Z}_{\odot} = 0, 10^{-4}, 10^{-3}$, and $10^{-2}$.
The gas is characterized by an initial temperature of $T_1 = 1.2 \times 10^4\ {\rm K}$ and 
density of $n_{\rm H, 1} = 4 \times 10^3\ {\rm cm}^{-3}$.
No UV irradiation is considered~($J_{21} = 0$).
In each panel, solid blue and red lines are the thermal tracks of clouds with and without dust, respectively,
and open circles and triangles correspond to the epochs of clump formation and its re-fragmentation.
The black-dashed lines are those of a pre-ionized cloud collapsing by self-gravity, not by shock compression, 
with the same initial temperature, metallicity, and dust content. 
The gray-shaded regions are temperatures below the CMB floor that is set at $T_{\rm CMB} = 30 \ {\rm K}$.
Diagonal dotted lines are the loci of constant Jeans mass.}
\label{fig:thermal_gasflow}
\end{figure*}

\begin{figure*}
\centering
\begin{tabular}{c}
{\includegraphics[scale=1.15]{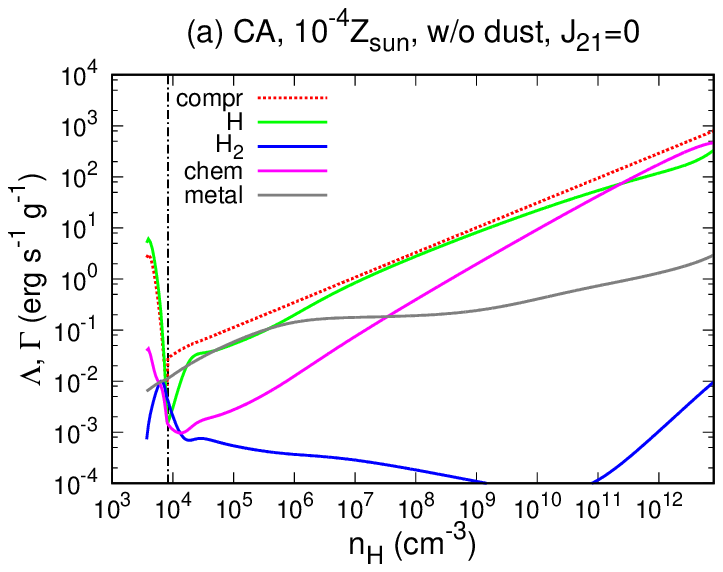}}
{\includegraphics[scale=1.15]{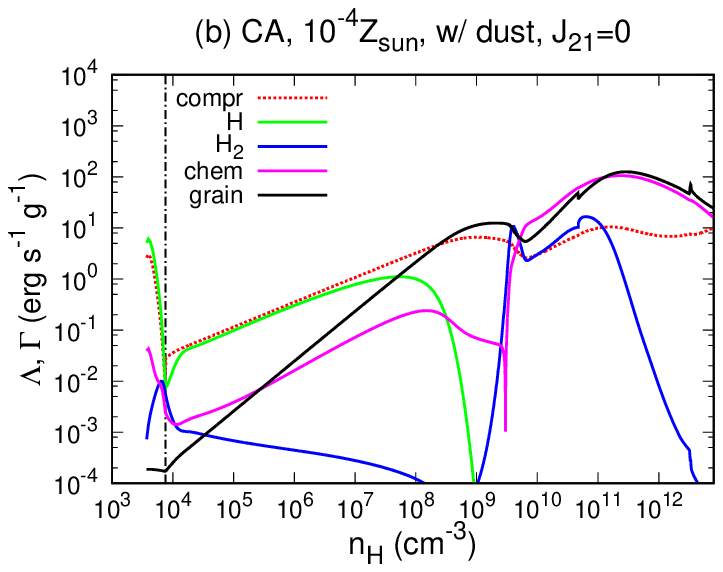}} \\
{\includegraphics[scale=1.15]{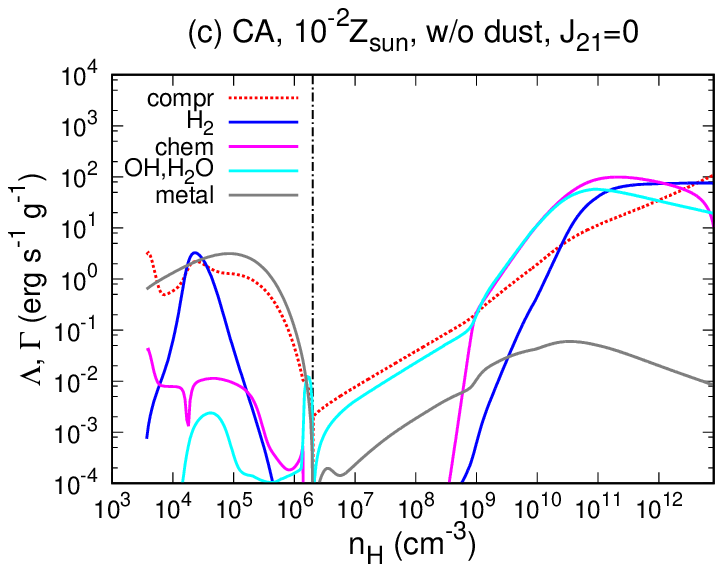}}
{\includegraphics[scale=1.15]{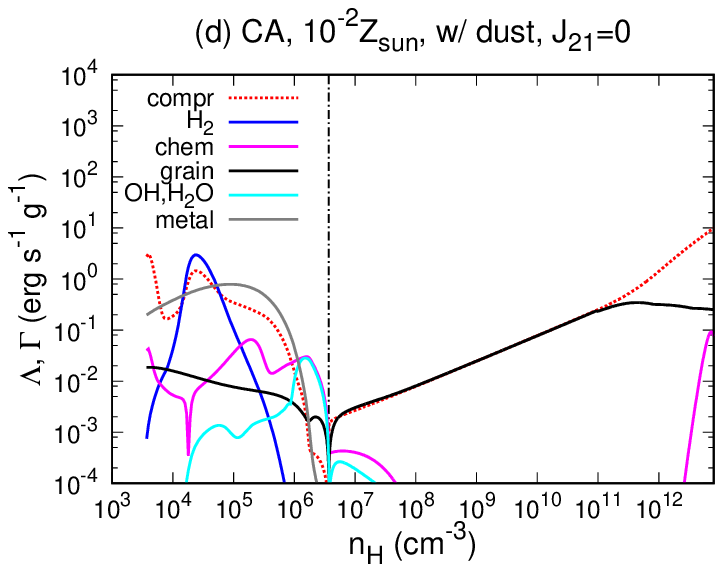}} \\
\end{tabular}
\caption{Cooling and heating rates in a cloud compressed by a CA shock.
Panels (a), (b), (c), and (d) show the models of ($10^{-4}~{\rm Z}_\odot$, without dust), ($10^{-4}~{\rm Z}_\odot$, with dust),
($10^{-2}~{\rm Z}_\odot$, without dust), and ($10^{-2}~{\rm Z}_\odot$, with dust), respectively.
Individual cooling/heating processes are indicated in the legend.
The vertical dot-dashed line indicates the epoch of clump formation.}
\label{fig:coolheat_gasflow}
\end{figure*}

Thermal evolution of clouds compressed by CA shocks in the absence of UV irradiation~($J_{21} = 0$) is shown
in Fig.~\ref{fig:thermal_gasflow}, for initial metallicities of
(a) $Z/{\rm Z}_{\sun}=0$, (b) $10^{-4}$, (c) $10^{-3}$, and (d) $10^{-2}$.
The blue and red lines in each panel show the cases with and without dust, respectively.
The circles and triangles on the thermal tracks indicate the epochs of clump formation and its re-fragmentation, respectively.
The black-dashed lines show the thermal tracks of initially ionized clouds~(`pre-ionized' clouds, hereafter)
that start their collapse without shock-compression, but with high ionization degree and temperature, similar to a shock-compressed gas~\citep{Omukai2012}.
They are presented for comparison to see how thermal tracks differ between a gas with and without shock-compression.
For the pre-ionized clouds, calculation is started from $n_{\rm H} = 1.0\ {\rm cm}^{-3}$, $T = 10^4\ {\rm K}$, and $y(e) = 0.1$
and the remaining conditions are the same as those of shocked clouds at each metallicity.
Shaded regions are temperatures below the CMB value, $T_{\rm CMB} = 30\ {\rm K}$, 
and the diagonal dotted lines show lines of constant Jeans masses.

First, we see the case of zero-metallicity gas~(Fig. \ref{fig:thermal_gasflow}a).
The isobaric contraction from the initial temperature and density of
$(T_1, n_{\rm H, 1}) = (1.2 \times 10^4{\rm K},\ 4 \times 10^3\ {\rm cm}^{-3})$ soon 
terminates at $(T, n_{\rm H}) = (8000{\rm K},\ 10^4\ {\rm cm}^{-3})$ since Ly$\alpha$ cooling has a cutoff at 8000K
and the H$_2$ formation is suppressed via collisional dissociation due to the high temperature and density~\citep{Inayoshi2012}.
At this moment, very massive clumps of $\sim 10^5\mbox{-}10^6\ {\rm M}_\odot$ form~(open red circle).
Subsequently, the clumps continue collapsing almost isothermally solely via Ly$\alpha$ cooling, 
with H$_2$ cooling being suppressed.
The clumps do not fragment again and supermassive stars of $\sim 10^5\mbox{-}10^6\ {\rm M}_\odot$ 
are expected to form~\citep[e.g.,][]{Nakauchi2017}, eventually leading to a black hole of comparable mass~\citep{Inayoshi_Omukai2014,Sugimura2017}.

The effects of increasing the gas metallicity are shown in the other panels of Fig. \ref{fig:thermal_gasflow}.
To interpret the evolutionary tracks, it is helpful to look at the relative importance 
of different cooling/heating processes, which are shown in Fig.~\ref{fig:coolheat_gasflow}
for the $10^{-4}~{\rm Z}_\odot$ and $10^{-2}~{\rm Z}_\odot$ models, with and without dust.

We first discuss the models without dust~(red solid lines in Fig.~\ref{fig:thermal_gasflow}).
The thermal track of $Z = 10^{-4}\ {\rm Z}_\odot$~(Fig.~\ref{fig:thermal_gasflow}b) is identical to the $Z=0$ track
since Ly$\alpha$ cooling is more effective than metal-line cooling~(Fig.~\ref{fig:coolheat_gasflow}a). 
The situation changes when the metallicity exceeds $Z \sim 10^{-3}\ {\rm Z}_\odot$~(Fig.~\ref{fig:thermal_gasflow}c, d).
The isobaric temperature drop continues well below 8000 K via metal-line cooling, until reaching the CMB floor. 
H$_2$~(H$_2$O) temporarily becomes the main coolant at $\simeq 1000$ K~(below $\simeq 50$ K, 
respectively, see Fig. \ref{fig:coolheat_gasflow}c).
Being unable to cool below the CMB temperature, 
the fragmentation condition~(Eq. \ref{eq:frag_cond}) is met at this stage,
and clumps of $\simeq 10\ {\rm M}_\odot$ form~(open red circles).
During the subsequent collapse of the clumps, the temperature increases rapidly both in 
$Z = 10^{-3}$ and $10^{-2}\ {\rm Z}_\odot$ cases, 
as radiative cooling, mostly by H$_2$O, cannot catch up with increasing compressional heating. 
When the density reaches $n_{\rm H} \sim 10^{9}-10^{10}\ {\rm cm}^{-3}$, H$_2$ formation via 
three-body reactions quickly proceeds and the associated heating balances H$_2$O cooling.
By $n_{\rm H} \sim 10^{11}\ {\rm cm}^{-3}$, all the hydrogen becomes molecular 
and H$_2$ formation heating declines thereafter.
Almost simultaneously, H$_2$ cooling becomes dominant, leading to a shallower temperature increase.
When $Z = 10^{-3}$ and $10^{-2}\ {\rm Z}_\odot$, the clump mass at the end of the isobaric evolution is always larger than
the instantaneous Jeans mass during the collapse phase and no further fragmentation occurs.
Therefore, the final clump mass is $M_{\rm clump} \simeq 10\ {\rm M}_\odot$.

Finally, we discuss the effects of dust~(blue solid lines in Fig.~\ref{fig:thermal_gasflow} and
Fig.~\ref{fig:coolheat_gasflow}b, d).
Dust affects the thermal evolution by means of (i) cooling by thermal emission,
(ii) H$_2$ formation on its surface, and when UV radiation is present, (iii) heating by photoelectric emission. 
The presence of dust has only a minor effect in the isobaric phase, 
although H$_2$ formation on grain surfaces and 
the associated heating become effective for the $Z/{\rm Z}_\odot = 10^{-3}$ and $10^{-2}$ cases.
When $Z = 10^{-2}\ {\rm Z}_\odot$, H$_2$ formation heating creates
a small bump in the thermal track at $T \simeq 40\ {\rm K}$, 
which makes the density at fragmentation slightly higher  
and thus the clump mass slightly smaller~($\lesssim 3\ {\rm M}_\odot$) compared to the models without dust.
In contrast, dust grains are an important coolant via the thermal emission in the subsequent collapse phase, 
leading to a large difference between the thermal 
tracks with and without dust (blue and red lines in Fig.~\ref{fig:thermal_gasflow}b-d).
Even when metallicity is as low as $Z = 10^{-4}\ {\rm Z}_\odot$~(Figs. \ref{fig:thermal_gasflow}b and \ref{fig:coolheat_gasflow}b), 
after the isothermal contraction at $\sim 8000$~K, the temperature suddenly drops to $\sim 800$~K at $n_{\rm H} \sim 10^9\ {\rm cm}^{-3}$ 
due to dust cooling.
Following the short interval of slight temperature increase at $n_{\rm H} \sim 10^{10}\mbox{-}10^{11}\ {\rm cm}^{-3}$ caused by 
H$_2$ formation heating, the temperature further drops to $\sim 100$~K due to dust cooling 
when H$_2$ formation is almost over.
Two open triangles at the local temperature minima indicate the two epochs of re-fragmentation.
The second fragmentation episode at $\sim 10^{13}\ {\rm cm^{-3}}$ produces very low-mass fragments of 
$M_{\rm re-frag} \sim 10^{-2}\ {\rm M}_\odot$.
When $Z = 10^{-3}\ {\rm Z}_{\odot}$~(Fig. \ref{fig:thermal_gasflow}c), after clump formation at $\sim 10^{6}\ {\rm cm}^{-3}$,
the temperature is raised first gradually and then abruptly at $\sim 10^{8}\ {\rm cm}^{-3}$ by H$_2$ formation heating via grain surface reactions.
The rapid heating at $\sim 10^{8}\ {\rm cm}^{-3}$ is caused by the increase in the fraction of H$_2$ formation energy released as heat at this density.
When H$_2$ formation is almost completed, 
the temperature drops suddenly to the CMB floor by dust cooling at $\sim 10^{10}\ {\rm cm}^{-3}$, 
resulting in the formation of sub-solar mass~($M_{\rm re-frag} \simeq 0.1\ {\rm M}_\odot$) fragments~(open triangle).
When $Z = 10^{-2}\ {\rm Z}_{\odot}$~(Figs. \ref{fig:thermal_gasflow}d and \ref{fig:coolheat_gasflow}d),
a clump formed at the end of the isobaric phase~(at $\sim 3 \times 10^{6}\ {\rm cm}^{-3}$) collapses
isothermally along the CMB floor thanks to efficient dust cooling, and no further fragmentation occurs.
Therefore, the clump mass does not change from the initial value~($M_{\rm clump} \lesssim 3\ {\rm M}_\odot$).
In all the models with dust, the clouds become optically thick to dust thermal emission at $n_{\rm H} \sim 10^{12}\mbox{-}10^{13}\ {\rm cm}^{-3}$ 
and contract adiabatically thereafter.

Black-dashed lines in Fig.~\ref{fig:thermal_gasflow} show that the pre-ionized clouds collapsing by self-gravity 
initially follow a very different evolution compared to the shock-compressed clouds.
The temperature is first lowered isochorically until $\simeq 200$~K via efficient H$_2$ cooling,
and then is pushed down to the CMB value at $n_{\rm H} \sim 10^3\mbox{-}10^4\ {\rm cm}^{-3}$,
via HD cooling for $Z \lesssim 10^{-4}\ {\rm Z}_\odot$ and metal-line cooling for $Z \gtrsim 10^{-3}\ {\rm Z}_\odot$.
After the temperature minima, H$_2$ formation heating via grain surface reactions increases the temperature,
until H$_2$ formation is almost completed and dust cooling becomes effective,
and the thermal tracks eventually converge to those of the shock-compressed cases.
When $Z \gtrsim 10^{-3}\ {\rm Z}_\odot$~(Fig.~\ref{fig:thermal_gasflow}c, d), this occurs at a density of 
$n_{\rm H} \sim 10^7\mbox{-}10^8\ {\rm cm}^{-3}$,
which is close to the epoch of clump formation in a shock-compressed cloud.
When $Z = 10^{-4}\ {\rm Z}_\odot$~(Fig.~\ref{fig:thermal_gasflow}b), the thermal tracks converge only after
the temperature of a clump is lowered via dust cooling from 8000 K to $\sim 300$~K at $n_{\rm H} \sim 10^{12}\ {\rm cm}^{-3}$.

\subsection{Cloud Compressed by a Supernova Shock}
\label{subsec:snremnant}

\begin{figure*}
\centering
\begin{tabular}{c}
{\includegraphics[scale=1.15]{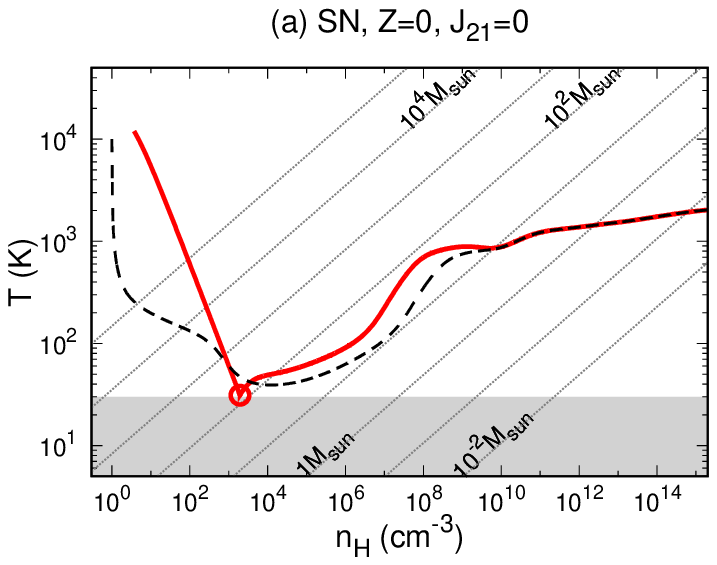}} 
{\includegraphics[scale=1.15]{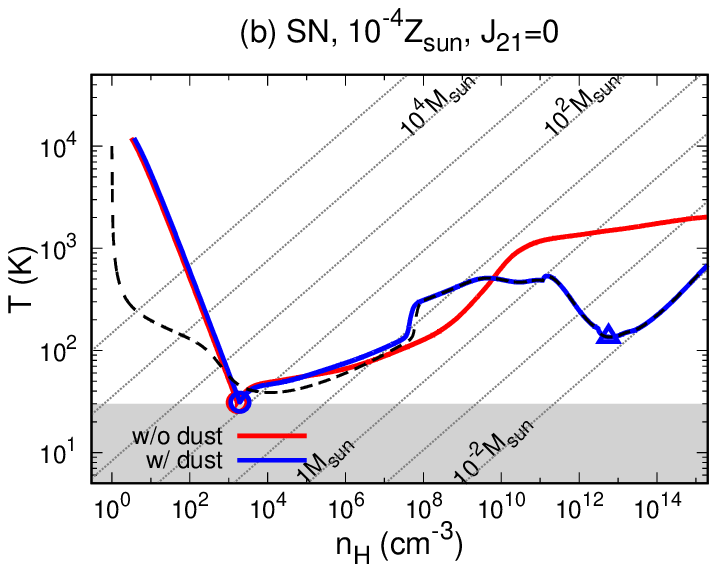}} \\
{\includegraphics[scale=1.15]{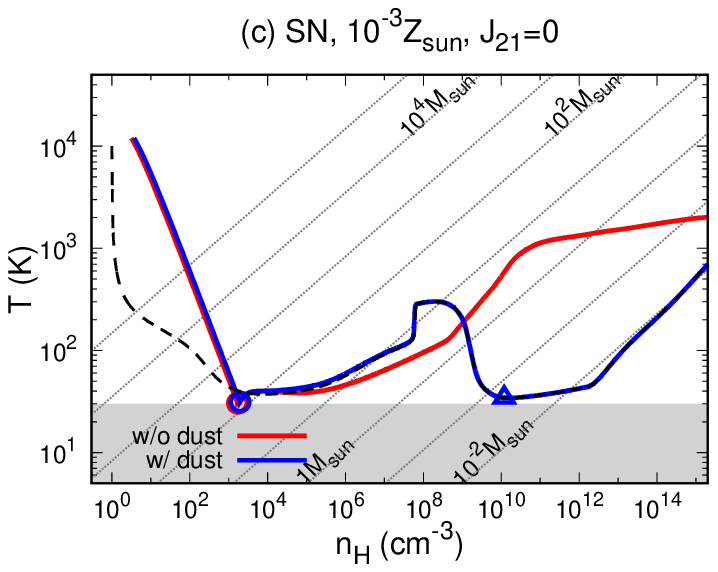}} 
{\includegraphics[scale=1.15]{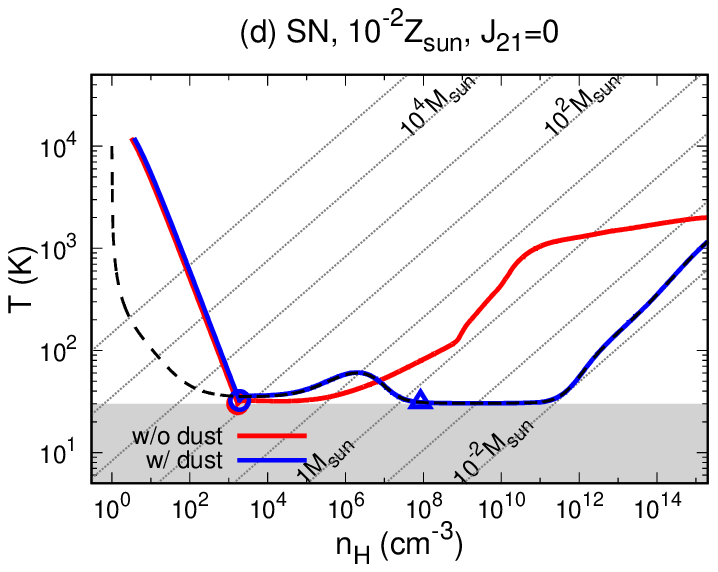}} \\
\end{tabular}
\caption{Same as Fig.~\ref{fig:thermal_gasflow}, but for a cloud compressed by a SN shock,
which is characterized by the initial density of $n_{\rm H, 1} = 4$~cm$^{-3}$.}
\label{fig:thermal_snshock}
\end{figure*}

\begin{figure*}
\centering
\begin{tabular}{c}
{\includegraphics[scale=1.15]{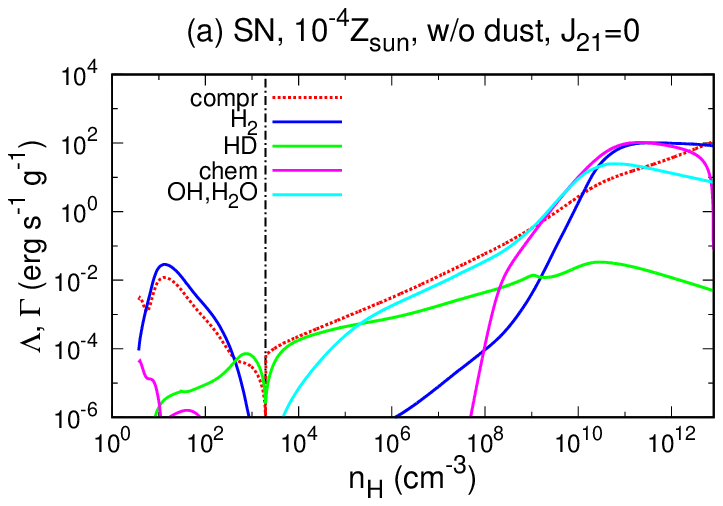}}
{\includegraphics[scale=1.15]{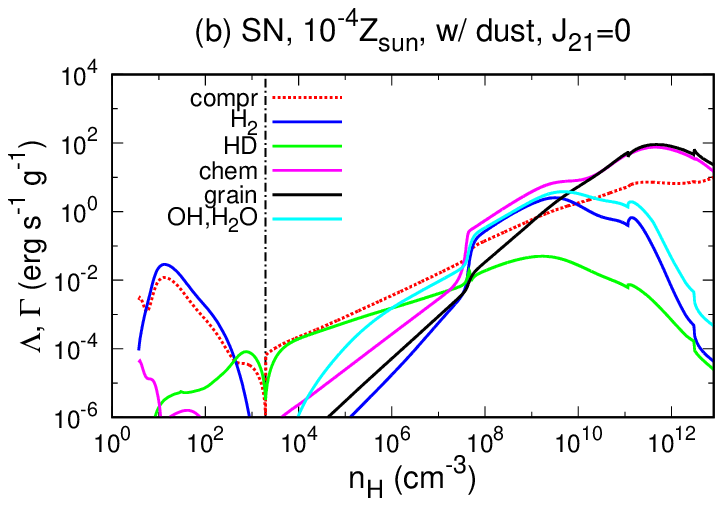}} \\
{\includegraphics[scale=1.15]{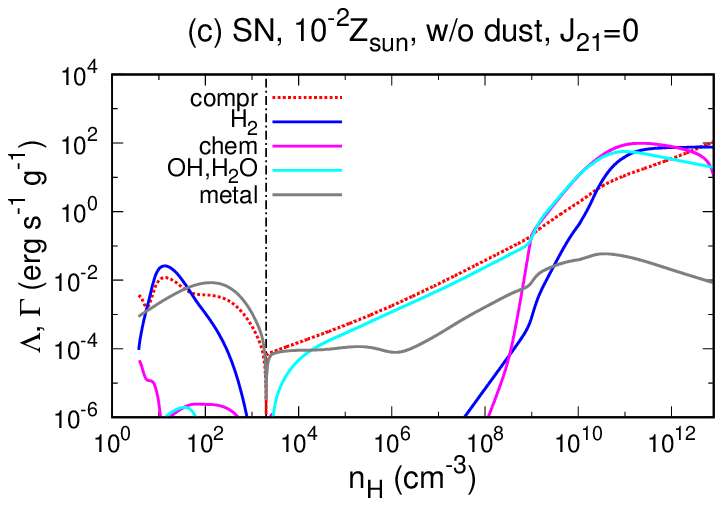}}
{\includegraphics[scale=1.15]{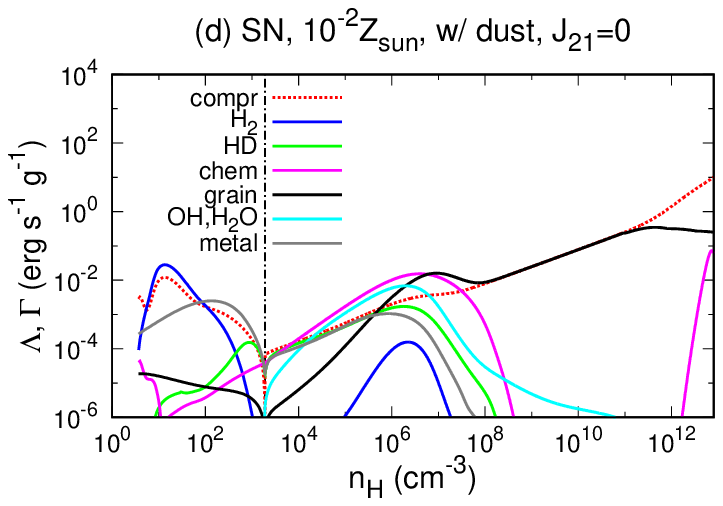}} \\
\end{tabular}
\caption{Same as Fig.~\ref{fig:coolheat_gasflow}, but for a cloud compressed by a SN shock,
which is characterized by the initial density of $n_{\rm H, 1} = 4$~cm$^{-3}$.}
\label{fig:coolheat_snshock}
\end{figure*}

The thermal evolution and the corresponding cooling/heating rates for a cloud compressed by a SN shock
are shown in Figs.~\ref{fig:thermal_snshock} and \ref{fig:coolheat_snshock}, respectively.
The main difference from the results of the CA shock is that 
the temperature reaches the CMB value at the end of the isobaric evolution
in all the models regardless of metallicity or presence of dust.
This is because when the initial density is $n_{\rm H, 1} \leq 10^3\ {\rm cm}^{-3}$, as in the SN case,
H$_2$ collisional dissociation is inefficient,
and the gas can first cool well below 8000 K via H$_2$ cooling.
When $Z \leq 10^{-4}\ {\rm Z}_\odot$, HD cooling takes over below 150 K until the temperature hits the CMB floor~(Fig.~\ref{fig:coolheat_snshock}a, b).
Metals also become important coolants during the isobaric contraction when $Z \geq 10^{-3}\ {\rm Z}_\odot$~(Fig.~\ref{fig:coolheat_snshock}c, d).
At $T \sim T_{\rm CMB}$, the fragmentation condition~(Eq. \ref{eq:frag_cond}) is satisfied
and clumps of $\sim 100\ {\rm M}_\odot$ are formed in all the models~(open circles in Fig.~\ref{fig:thermal_snshock}).

The subsequent clump evolution depends strongly on whether dust is present or not. 
Without dust~(solid red lines in Fig.~\ref{fig:thermal_snshock}), the temperature continues
to increase first by compressional heating~($\lesssim 10^{8}\ {\rm cm^{-3}}$)
and then by three-body H$_2$ formation heating at higher densities~(Fig.~\ref{fig:coolheat_snshock}a, c).
The main cooling channels are provided by molecular lines of H$_2$, HD, OH and H$_2$O 
and for $Z = 10^{-2}\ {\rm Z}_\odot$ also by fine-structure lines,
but they fail to trigger another phase of rapid cooling and fragmentation.
The final clump mass is thus set by the value at the end of the isobaric phase as $M_{\rm clump} \simeq 100\ {\rm M}_\odot$.
In the presence of dust grains~(solid blue lines in Fig.~\ref{fig:thermal_snshock})
the thermal evolution after clump formation at $\sim 10^3\ {\rm cm}^{-3}$ becomes very similar to the pre-ionized cases~(black-dashed lines),
despite the very different behaviors below that density.
During the collapse, the temperature is raised by H$_2$ formation heating via grain surface reactions,
and is then lowered rapidly to the local minima through dust cooling at $n_{\rm H} \sim 10^8\mbox{-}10^{12}\ {\rm cm}^{-3}$~(Fig.~\ref{fig:coolheat_snshock}b, d), where sub-solar mass fragments~($M_{\rm re-frag} \lesssim 1\ {\rm M}_\odot$) are formed by re-fragmentation of the clumps~(open triangles in Fig.~\ref{fig:thermal_snshock}).

\subsection{Effect of UV Irradiation}\label{subsec:UVfield}

\begin{figure*}
\centering
\begin{tabular}{c}
{\includegraphics[scale=1.15]{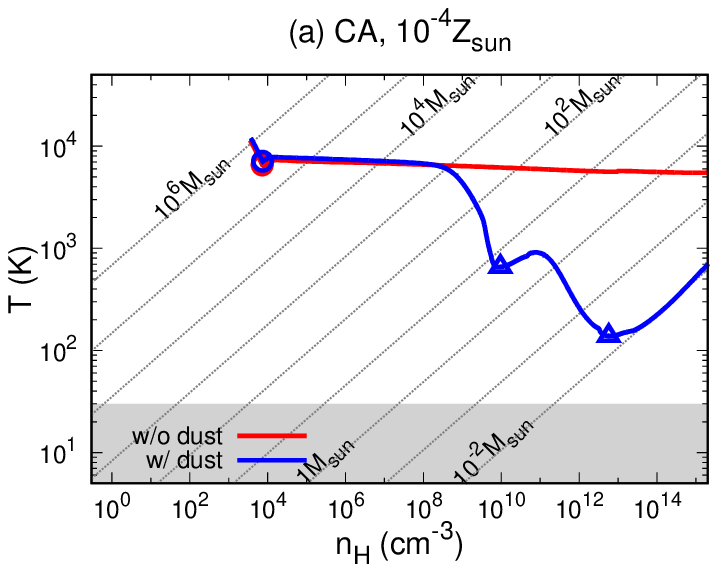}}
{\includegraphics[scale=1.15]{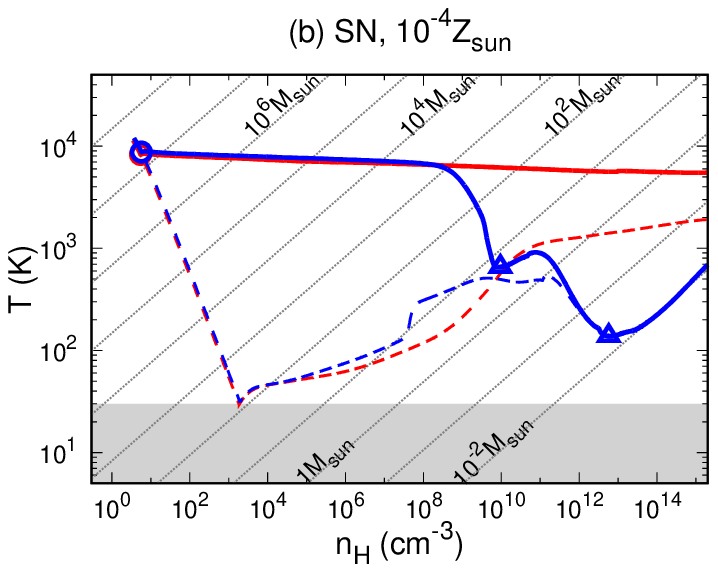}} \\
{\includegraphics[scale=1.15]{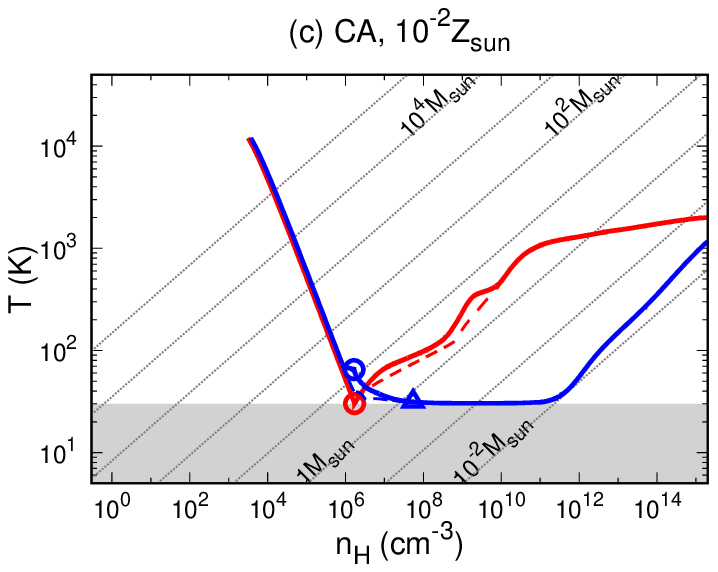}}
{\includegraphics[scale=1.15]{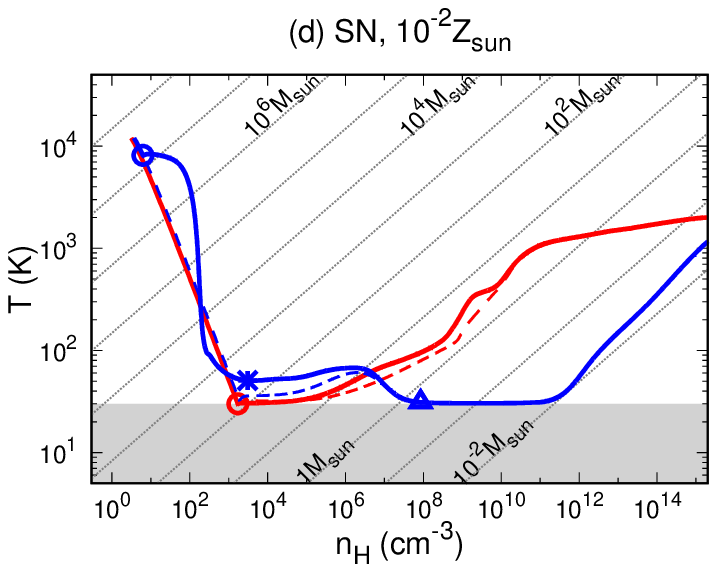}} \\
\end{tabular}
\caption{Same as Figs.~\ref{fig:thermal_gasflow} and~\ref{fig:thermal_snshock},
but showing the effects of UV irradiation on thermal evolution.
In each panel, solid and dashed lines show the results with $J_{21} = 10^4$ and $J_{21} = 0$, respectively.
Panels (a), (b), (c), and (d) refer to the models of (CA, $10^{-4}\ {\rm Z}_\odot$), (SN, $10^{-4}\ {\rm Z}_\odot$),
(CA, $10^{-2}\ {\rm Z}_\odot$), and (SN, $10^{-2}\ {\rm Z}_\odot$), respectively.
On thermal tracks, open circles indicate the epochs of clump formation and asterisks and open triangles indicate those of re-fragmentation.
}
\label{fig:thermal_uv}
\end{figure*}

Here, we discuss the effects of UV irradiation on the thermal evolution and fragmentation,
by comparing the results obtained assuming $J_{21} = 10^4$~(solid lines) and $J_{21} = 0$~(dashed lines) in Fig.~\ref{fig:thermal_uv}.
Individual panels refer to the CA models with $10^{-4}\ {\rm Z}_\odot$ and $10^{-2}\ {\rm Z}_\odot$~(panels a and c) and to the SN models with $10^{-4}\ {\rm Z}_\odot$ and $10^{-2}\ {\rm Z}_\odot$~(panels b and d).
Open circles indicate the epochs of clump formation, and asterisks and open triangles indicate those of re-fragmentation
caused by line and dust cooling, respectively.

UV irradiation has little impacts on the thermal path in the case of a CA shock~(Fig.~\ref{fig:thermal_uv}a, c).
With the high initial density of $n_{\rm H, 1} = 4 \times 10^3\ {\rm cm}^{-3}$, H$_2$ cooling is quenched
by collisional dissociation at a temperature higher than $\sim 8000\ {\rm K}$, independently of the presence of UV irradiation.
When $Z = 10^{-4}\ {\rm Z}_\odot$~(Fig.~\ref{fig:thermal_uv}a),
a very massive clump~($\sim 10^5\mbox{-}10^6\ {\rm M}_\odot$) forms at $(n_{\rm H}, T) \sim (10^4\ {\rm cm^{-3}}, 8000\ {\rm K})$~(open circles),
as a result of the isobaric evolution via Ly$\alpha$ cooling.
This clump follows the same evolutionary track of the $J_{21} = 0$ model discussed in Section \ref{subsec:gasflow}.
With more metals~($Z \gtrsim 10^{-3}\ {\rm Z}_\odot$, e.g., Fig.~\ref{fig:thermal_uv}c), efficient metal line-cooling decreases the gas temperature below 100 K in the isobaric phase independently of the UV strength.
In the model without dust~(red line), a clump is formed at $(n_{\rm H}, T) \sim (10^6\ {\rm cm^{-3}}, T_{\rm CMB})$~(open red circle).
Since H$_2$O is not shielded against UV in our model, H$_2$O formation remains suppressed via photodissociation 
for $n_{\rm H} \sim 10^6\mbox{-}10^{10}\ {\rm cm}^{-3}$ and the clump collapses 
with a slightly higher temperature compared to the $J_{21}=0$ case. 
At $n_{\rm H} > 10^{10}\ {\rm cm}^{-3}$, H$_2$O formation overcomes photodissociation and 
the clump follows the same path as the $J_{21}=0$ case.
In the model with dust~(blue line), photoelectric heating cannot be neglected at $n_{\rm H} \sim 10^5\mbox{-}10^6\ {\rm cm}^{-3}$, 
where UV is still not shielded effectively by dust absorption, and the cooling rate becomes less efficient, increasing the cooling time.
As a result, the gas satisfies the fragmentation condition~(Eq. \ref{eq:frag_cond}) at a temperature~($\sim 100$~K; open blue circle)
higher than in the $J_{21} = 0$ case, which also makes the clump mass~($M_{\rm clump} \simeq 20\ {\rm M}_\odot$) several times larger.
Soon after clump formation, the temperature drops to the CMB value via  dust thermal emission
and the clump experiences another episode of fragmentation into solar-mass~($M_{\rm re-frag} \simeq 1\ {\rm M}_\odot$) pieces~(open triangle).
Although this re-fragmentation is lacking in the $J_{21} = 0$ model, 
the evolutionary tracks of models with $J_{21} = 0$ and $10^{4}$ converge thereafter.

A cloud compressed by a SN shock is far more vulnerable to UV irradiation, 
owing to the lower initial density and weaker UV shielding~(Fig.~\ref{fig:thermal_uv}b, d).
When H$_2$ cooling is strongly suppressed by photodissoiciation,
the gas can cool isobarically only up to 8000 K via Ly$\alpha$ cooling, for most of the models.
In these cases, the clump mass becomes very large $M_{\rm clump} \sim 10^7\ {\rm M}_\odot$,
due to the very low density $\sim 10\ {\rm cm}^{-3}$ and high temperature of fragmentation~(open circles).
When $Z = 10^{-4}\ {\rm Z}_\odot$~(Fig.~\ref{fig:thermal_uv}b),
the evolutionary tracks after clump formation are almost the same as in the CA case~(Fig.~\ref{fig:thermal_uv}a).
In the model with dust and with $Z = 10^{-2}\ {\rm Z}_\odot$~(the blue line in Fig.~\ref{fig:thermal_uv}d),
the temperature in a collapsing clump drops isochorically at $n_{\rm H} \sim 10^2\ {\rm cm}^{-3}$ via metal-line cooling.
When the temperature reaches $\sim 60$~K, metal-line cooling becomes inefficient and 
the clump fragments into massive cores of $M_{\rm re-frag} \sim 300\ {\rm M}_\odot$~(asterisk).
These fragments then contract almost isothermally until $n_{\rm H} \sim 10^{6}\ {\rm cm}^{-3}$, 
where the temperature again starts decreasing via dust cooling.
The temperature is lowered to the CMB value at $n_{\rm H} \sim 10^{8}\ {\rm cm}^{-3}$, where
further fragmentation into much smaller pieces of $M_{\rm re-frag} \simeq 1\ {\rm M}_\odot$ occurs~(open triangle).
Thereafter, the evolutionary track converges on the $J_{21} =0$ case.
Note that in the model without dust and with $Z = 10^{-2}\ {\rm Z}_\odot$~(the red line in Fig.~\ref{fig:thermal_uv}d),
the temperature reaches the CMB value during the isobaric evolution, even in the $J_{21} =10^4$ model.
This is because, for a fixed metallicity, a model without dust contains a larger amount of metals in the gas-phase
and efficient metal-line cooling decreases the temperature below 8000 K.
During the collapse, the thermal track becomes identical with that in the CA case at $n_{\rm H} > 10^{7}\ {\rm cm}^{-3}$.

\subsection{Effect of CMB}\label{subsec:CMBfield}

\begin{figure*}
\centering
\begin{tabular}{c}
{\includegraphics[scale=1.15]{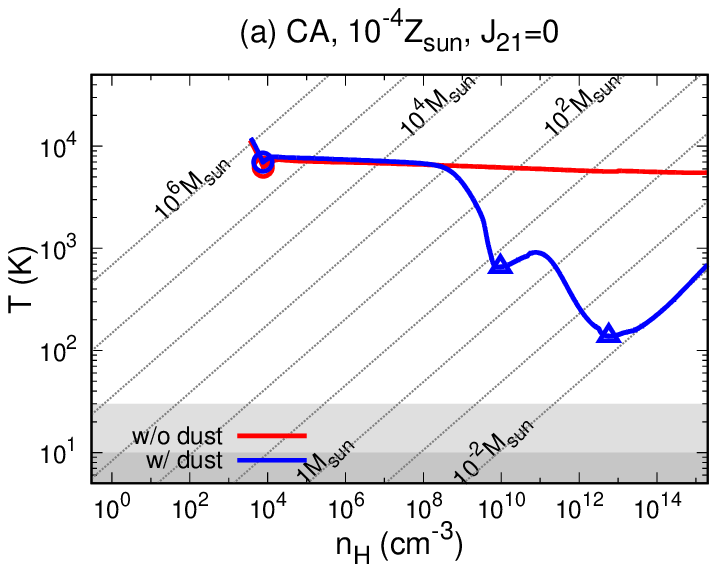}}
{\includegraphics[scale=1.15]{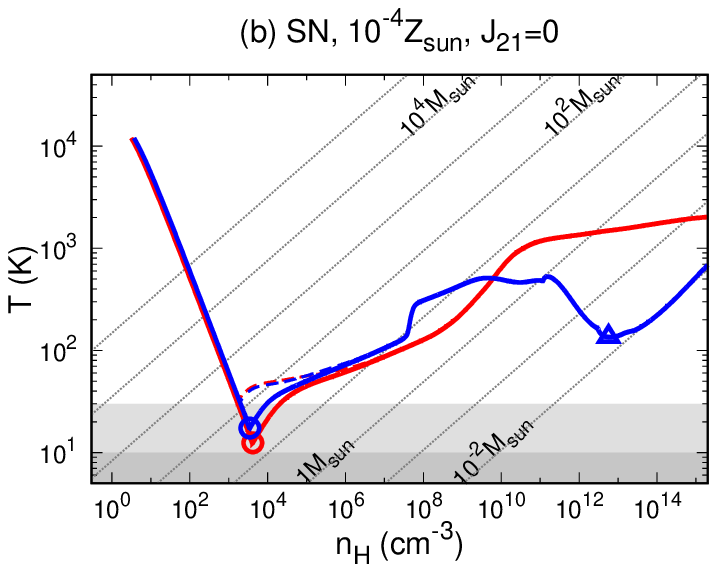}} \\
{\includegraphics[scale=1.15]{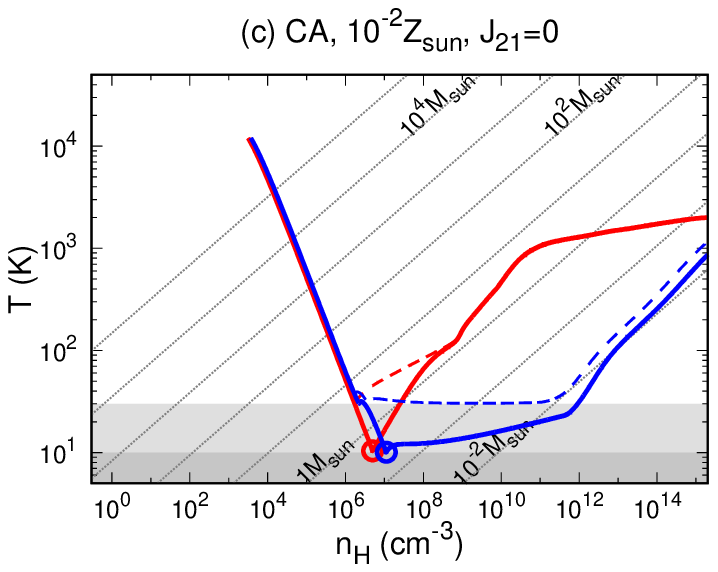}}
{\includegraphics[scale=1.15]{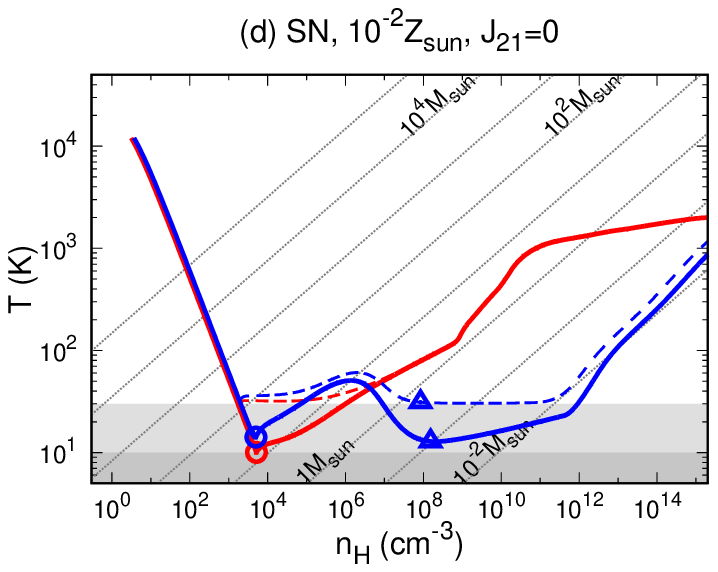}} \\
\end{tabular}
\caption{Same as Fig.~\ref{fig:thermal_uv}, but showing the CMB effects.
Solid and dashed lines show the results with $T_{\rm CMB} = 10\ {\rm K}$ and $T_{\rm CMB} = 30\ {\rm K}$, respectively.
}
\label{fig:thermal_cmb}
\end{figure*}

The CMB affects the low-metallicity gas evolution and fragmentation by setting a minimum temperature floor that gas can reach,
as studied in \cite{Schneider2010} by a semi-analytical approach and in \cite{Smith2009} by three-dimensional hydrodynamic calculations, although the latter does not take dust cooling and H$_2$ formation heating into account.
Here, we see the CMB effects by changing the CMB temperature to $T_{\rm CMB} = 10\ {\rm K}$.
In Fig. \ref{fig:thermal_cmb}, we show the thermal tracks with $T_{\rm CMB} = 10\ {\rm K}$~(solid lines) 
for the CA models with $10^{-4}\ {\rm Z}_\odot$ and $10^{-2}\ {\rm Z}_\odot$~(panels a and c) and the SN models with $10^{-4}\ {\rm Z}_\odot$ and $10^{-2}\ {\rm Z}_\odot$~(panels b and d). 
The models with $T_{\rm CMB} = 30\ {\rm K}$ are also shown for comparison by dashed lines.

When efficient coolants are available, 
a shocked layer cools isobarically until the temperature hits the CMB floor at 10 K, 
where clumps are formed by fragmentation.
This is the case when $Z \gtrsim 10^{-3}\ {\rm Z}_\odot$ in CA shocks, 
and the clump mass is $\sim 0.3\ {\rm M}_\odot$ both in models with and without dust~(Fig. \ref{fig:thermal_cmb}a, c).
In SN shocks, the gas reaches the CMB temperature floor at $10\ {\rm K}$ in all the models shown in Fig. \ref{fig:thermal_cmb}b, d 
and clumps with masses $\sim 10\mbox{-}30\ {\rm M}_\odot$ form.
Thanks to the lower value of $T_{\rm CMB}$, the clump mass is one order of magnitude smaller than the $T_{\rm CMB}=30\ {\rm K}$ model.
Note that the clump mass at the end of the isobaric evolution can be inferred from the isobaric relation, $n_{\rm H,1} T_1 \simeq n_{\rm frag} T_{\rm CMB}$, as 
\begin{equation}
\begin{split}
M_{\rm J}(n_{\rm frag},T_{\rm CMB}) &\sim 0.3\ {\rm M}_\odot \left(\frac{T_{\rm CMB}}{10\ {\rm K}} \right)^2\left(\frac{\mu}{2} \right)^{-2} \\
&\quad \times \left(\frac{n_{\rm H,1}}{4000\ {\rm cm}^{-3}} \frac{T_1}{12000\ {\rm K}} \right)^{-1/2}
\end{split}
\label{eq:jeans_mass2}
\end{equation}
and it becomes $\lesssim 3\ {\rm M}_\odot$
if the initial density is $n_{\rm H,1} \gtrsim 40\ {\rm cm}^{-3}~(T_{\rm CMB}/10\ {\rm K})^4$.

In the collapse phase of a clump, CMB effect appears explicitly in the models with dust and with $Z \gtrsim 10^{-3}\ {\rm Z}_\odot$~(
e.g., blue lines in Fig. \ref{fig:thermal_cmb}c, d).
In fact, the temperature remains below $30\ {\rm K}$
until the clump becomes optically thick to dust emission at $n_{\rm H} \sim 10^{12}\ {\rm cm^{-3}}$.
In particular, in the SN shock case, dust cooling triggers the re-fragmentation of a clump
into $\sim 0.1\ {\rm M}_{\sun}$ fragments at $n_{\rm H} \sim 10^{8}\ {\rm cm^{-3}}$~(open triangle in Fig. \ref{fig:thermal_cmb}d).
This mass scale decreases as the CMB temperature decreases.
In the other models, however, the temperature increases above 30 K and 
the thermal tracks soon converge with those of $T_{\rm CMB} = 30\ {\rm K}$ models~(cf. red and blue lines in Fig. \ref{fig:thermal_cmb}).
CMB effect is hardly observed in these cases.
Note that in the CA model without dust and with $Z = 10^{-2}\ {\rm Z}_\odot$~(the red line in Fig. \ref{fig:thermal_cmb}c),
owing to the rapid temperature increase at $n_{\rm H} \sim 10^7\mbox{-}10^{10}\ {\rm cm}^{-3}$,
the clump can collapse by accreting materials and elevating its mass from $\sim 0.3\ {\rm M}_\odot$ to $\sim 10\ {\rm M}_\odot$, i.e.,
the Jeans mass at ($n_{\rm H}, T) \sim (10^{10}\ {\rm cm}^{-3}, 10^3\ {\rm K}$).
Therefore, the final clump mass is $M_{\rm clump} \sim 10\ {\rm M}_\odot$.

\section{Fragment Mass}\label{sec:fragment_mass}

Here we summarize the fragment mass scales in shock-compressed clouds, 
and discuss their dependence on the environmental conditions, 
i.e., initial metallicity, UV strength, and the presence of dust for 
both the CA and SN shocks.
The condition for low-mass star formation is summarized in Section \ref{subsec:condition} for a wider 
range of initial densities.

\subsection{Dependence of fragment mass on metallicity, UV strength, and presence of dust}
\label{subsec:fragment_mass}

\begin{figure*}
\centering
\begin{tabular}{c}
{\includegraphics[scale=1.15]{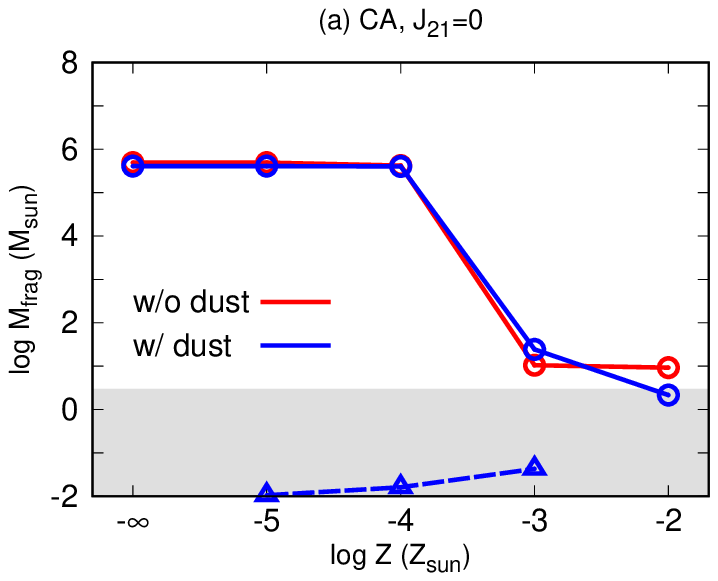}}
{\includegraphics[scale=1.15]{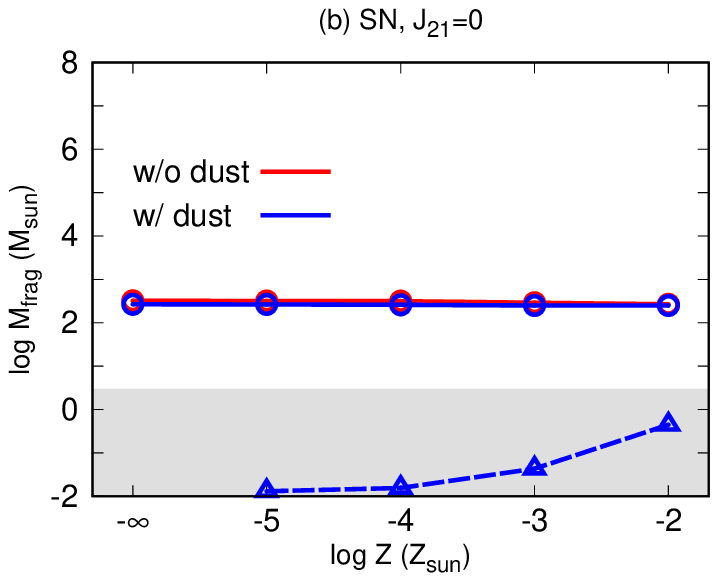}} \\
{\includegraphics[scale=1.15]{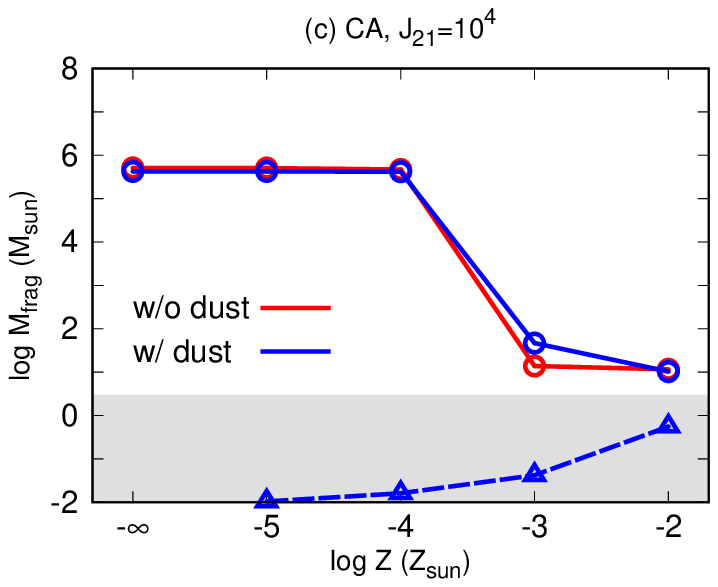}} 
{\includegraphics[scale=1.15]{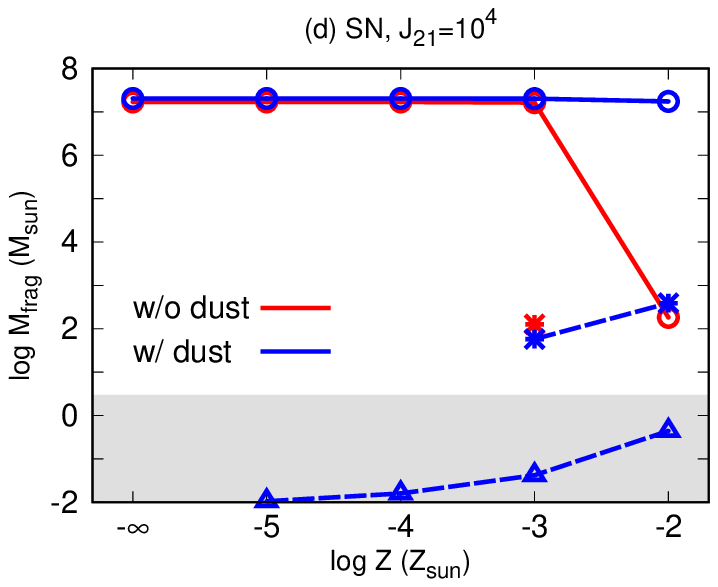}} \\
\end{tabular}
\caption{
Fragment mass as a function of metallicity, and its dependence on the initial density, UV strength, and the presence of dust.
Panels (a), (b), (c), and (d) refer to the models of (CA, $J_{21} = 0$), (SN, $J_{21} = 0$),
(CA, $J_{21} = 10^4$), and (SN, $J_{21} = 10^4$), respectively.
In each panel, blue and red colors indicate the clouds with and without dust.
Open circles connected by solid lines, and asterisks and open triangles connected by dashed lines
are used to identify the fragment mass scales of $M_{\rm clump}$, and $M_{\rm re-frag}$, respectively.
Sub-solar mass stars are expected to form in the gray-shaded regions~(see text).}
\label{fig:frag_masses}
\end{figure*}

In Fig.~\ref{fig:frag_masses}, the fragment mass scales are shown as a function of the initial metallicity
for CA shock models with $J_{21} = 0$ and $10^4$~(panels a and c), and SN shock models with $J_{21} = 0$ and $10^4$~(panels b and d).
In each panel, blue and red colors indicate the clouds with and without dust.
Open circles connected by solid lines indicate the clump mass $M_{\rm clump}$.
Asterisks and open triangles connected by dashed lines show the re-fragment mass scales $M_{\rm re-frag}$
determined by metal-line cooling and dust cooling, respectively.
The gray-shaded region in each panel marks the range of values where $M_{\rm frag} \leq 3\ {\rm M}_\odot$ and sub-solar mass stars form when the star formation efficiency is $\simeq 30\%$.
The star formation efficiency is observationally estimated to be in the range of $\simeq 20\mbox{-}40\%$,
by comparing the shape and amplitude of the prestellar core mass function 
with those of the stellar initial mass function~\citep[e.g.,][]{Alves2007, Andre2010}.
Using three-dimensional resistive magneto-hydrodynamical~(MHD) simulations, \cite{Machida2012} suggested that less than half of the mass of a prestellar cloud becomes a star and the rest is released into the interstellar space by a protostellar outflow.

We first discuss the results for CA shocks~(Fig.~\ref{fig:frag_masses}a, c).
Owing to the high initial density, 
the thermal evolution and thus the fragment mass scales~(both $M_{\rm clump}$ and $M_{\rm re-frag}$)
are hardly affected by UV irradiation for most of the models~(see also Section \ref{subsec:UVfield}).
The clump mass $M_{\rm clump}$~(open circle) is almost entirely determined by the gas metallicity:
it is $\sim 10^6\ {\rm M}_{\odot}$ for metallicities lower than $\sim 10^{-4}\ {\rm Z}_{\odot}$ and 
decreases abruptly to $\simeq 10\ {\rm M}_\odot$ for $Z \gtrsim 10^{-3}\ {\rm Z}_{\odot}$.
The presence of dust increases~(lowers) the clump mass $M_{\rm clump}$ only slightly for $Z = 10^{-3}\ {\rm Z}_{\odot}~(10^{-2}\ {\rm Z}_{\odot}$, respectively).
In models with dust~(open triangles), these clumps re-fragment into sub-solar mass pieces 
and low-mass star formation becomes possible when $10^{-5} \lesssim Z/{\rm Z}_{\odot} \lesssim 10^{-3}$ for the $J_{21} = 0$ case
and $Z/{\rm Z}_{\odot} \gtrsim 10^{-5}$ for the $J_{21} = 10^4$ case, respectively.

In SN shock models~(Fig.~\ref{fig:frag_masses}b, d),
due to the low initial density, the thermal evolution and fragment mass scale are greatly affected by UV irradiation.
In the absence of UV irradiation~(Fig.~\ref{fig:frag_masses}b, open circles), the clump mass is $M_{\rm clump} \simeq 200\ {\rm M}_\odot$ at all initial metallicities.
Under the strong UV field of $J_{21} = 10^4$~(Fig.~\ref{fig:frag_masses}d, open circles), 
the clump mass is boosted to $M_{\rm clump} \sim 10^7\ {\rm M}_{\odot}$.
The only exception is the model with $Z = 10^{-2}\ {\rm Z}_\odot$ and no dust,
where the gas can cool isobarically to $T_{\rm CMB}$ by metal-line cooling
and no re-fragmentation is expected thereafter,
so that the clump mass $M_{\rm clump}$ is the same as that in the $J_{21} = 0$ case.
In SN shock models with $Z \gtrsim 10^{-3}\ {\rm Z}_\odot$,
although metal-line cooling triggers re-fragmentation of clumps, the fragment mass 
$M_{\rm re-frag}$ is higher than $\sim 30\ {\rm M}_{\sun}$~(asterisks in Fig.~\ref{fig:frag_masses}d).
In the models with dust, these clumps fragment again at high enough densities~($\gtrsim 10^{8}\ {\rm cm^{-3}}$)
thanks to dust cooling and so the fragment mass $M_{\rm re-frag}$ can be very small~($\simeq 0.01\mbox{-}1\ {\rm M}_\odot$) 
when $Z \gtrsim 10^{-5}\ {\rm Z}_\odot$~(open triangles), comparable to CA shock cases. 

\subsection{Condition for low-mass star formation}~\label{subsec:condition}

\begin{figure*}
\centering
\begin{tabular}{c}
{\includegraphics[scale=1.15]{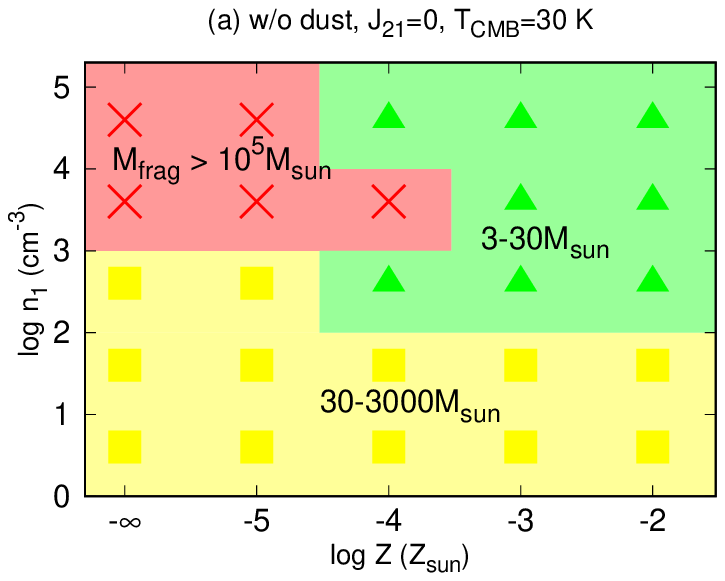}}
{\includegraphics[scale=1.15]{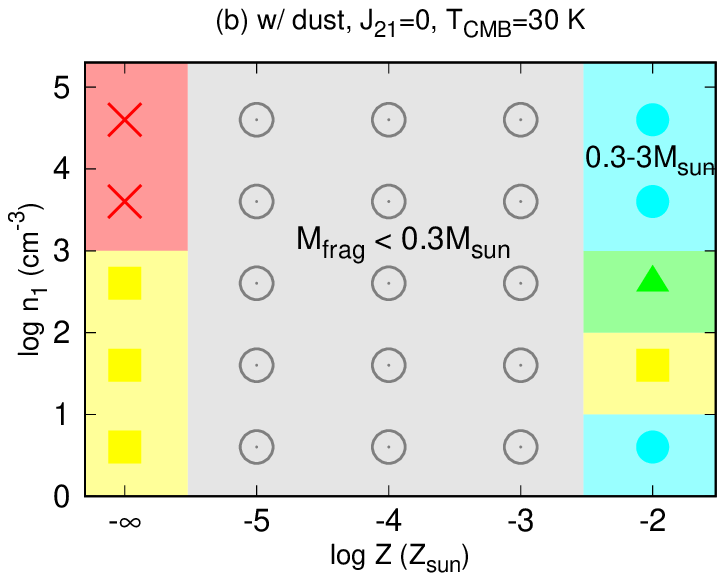}} \\
{\includegraphics[scale=1.15]{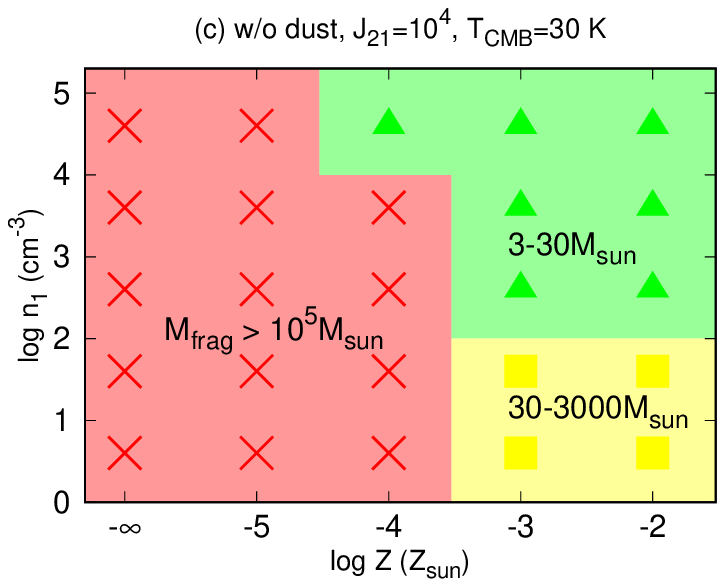}} 
{\includegraphics[scale=1.15]{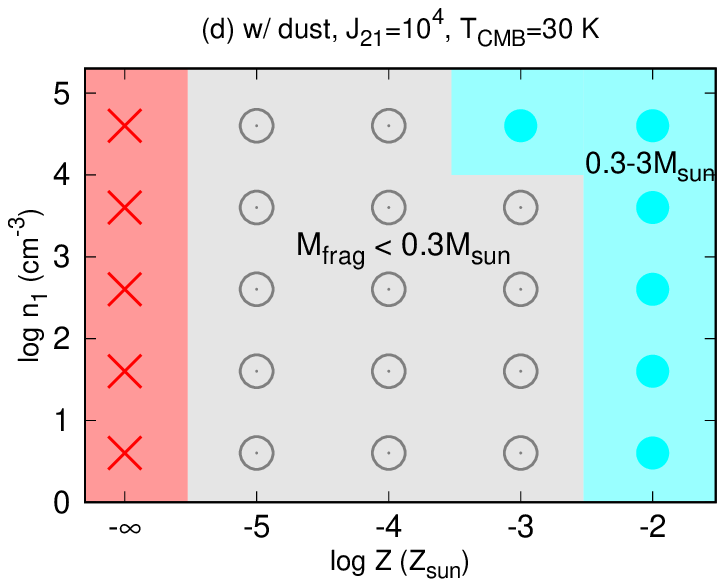}} \\
\end{tabular}
\caption{Conditions for the formation of 
very low mass~($M_{\rm frag} < 0.3\ {\rm M}_\odot$; grey shaded, open circles), 
low mass~($M_{\rm frag} = 0.3\mbox{-}3\ {\rm M}_\odot$; cyan shaded, filled circles), 
intermediate mass~($M_{\rm frag} = 3\mbox{-}30\ {\rm M}_\odot$; green shaded, triangles), 
massive~($M_{\rm frag} = 30\mbox{-}3000\ {\rm M}_\odot$; yellow shaded, squares), and 
very massive~($M_{\rm frag} > 10^5\ {\rm M}_\odot$; red shaded, crosses) fragments.
In this plot, $M_{\rm frag}$ means the minimum fragment mass during the evolution, 
i.e., $M_{\rm frag} = {\rm min}[M_{\rm clump}, M_{\rm re-frag}]$.
Panels (a) and (b) refer to models of $J_{21} = 0$ without and with dust, and
panels (c) and (d) refer to models of $J_{21} = 10^4$ without and with dust, respectively.
}
\label{fig:condition_Tr30}
\end{figure*}

Here we discuss the final fragment mass in the collapse phase, which is given by 
$M_{\rm frag} = {\rm min}(M_{\rm clump}, M_{\rm re-frag})$, for a wider range of initial densities 
$n_{\rm H, 1} = 1\mbox{-}10^{5}\ {\rm cm^{-3}}$ and metallicities $Z = 0\mbox{-}10^{-2}\ {\rm Z}_{\sun}$.
Fig. \ref{fig:condition_Tr30} shows models of $J_{21} = 0$ without and with dust~(panels a and b), and $J_{21} = 10^4$ without and with dust~(panels c and d).
Each panel is divided into five domains of ($Z$, $n_{\rm H, 1}$), depending on the value of $M_{\rm frag}$: 
(i) very low-mass~($M_{\rm frag} < 0.3\ {\rm M}_\odot$; grey shaded, open circles),
(ii) low-mass~($\sim 0.3-3\ {\rm M}_\odot$; cyan shaded, filled circles), 
(iii) intermediate mass~($\sim 3-30\ {\rm M}_\odot$; green shaded, triangles), 
(iv) massive~($\sim 30-3000\ {\rm M}_\odot$; yellow shaded, squares), and 
(v) very massive~($ > 10^5\ {\rm M}_\odot$; red shaded, crosses) fragments.
The CMB temperature is set at $T_{\rm CMB} = 30\ {\rm K}$.

In all the models without dust~(Fig. \ref{fig:condition_Tr30}a, c), the fragment mass is higher than $\sim 3\ {\rm M}_{\sun}$
and there is no parameter space forming low-mass fragments.
When $J_{21} = 0$~(Fig. \ref{fig:condition_Tr30}a) and $n_{\rm H, 1} \lesssim 100\ {\rm cm}^{-3}$,
massive fragments~(in the 30-3000 M$_\odot$ range) form, and this value depends on the initial metallicity only weakly.
In models with $n_{\rm H, 1} \gtrsim 10^3\ {\rm cm}^{-3}$, the fragment mass scale bifurcates into
the intermediate scale range~(3-30 M$_\odot$) for $Z \gtrsim 10^{-4}\ {\rm Z}_\odot$ and
the very massive scale range~($\gtrsim 10^5\ {\rm M}_\odot$) for $Z \lesssim 10^{-4}\ {\rm Z}_\odot$.
The UV irradiation greatly elevates the fragment mass scale from the massive to the very massive scale
in models with $Z \lesssim 10^{-4}\ {\rm Z}_\odot$ and $n_{\rm H, 1} \lesssim 10^3\ {\rm cm}^{-3}$, but
its effect is not observed in other models~(Fig. \ref{fig:condition_Tr30}c).
In the models with dust~(Fig. \ref{fig:condition_Tr30}b, d), as a result of clump re-fragmentation,
very low-mass~($< 0.3\ {\rm M}_\odot$) and low-mass~(0.3-3\ M$_\odot$) fragments are formed 
for $10^{-5} \lesssim Z/{\rm Z}_\odot \lesssim 10^{-3}$ and $Z/{\rm Z}_\odot = 10^{-2}$, respectively.
This trend is observed in almost all the models with dust.
In conclusion, the presence of dust is indispensable for the formation of sub-solar mass stars also in shock-compressed clouds.

\section{Summary and Discussion}
\label{sec:discussion}

We have studied the thermal evolution of shock-compressed clouds starting from 
a temperature of $\sim 10^4$ K, for a large range of metallicities $Z=0\mbox{-}10^{-2}\ {\rm Z}_{\sun}$,
post-shock densities $n_{\rm H,1} = 1\mbox{-}10^5\ {\rm cm^{-3}}$, and CMB temperatures $T_{\rm CMB}=10$ and $30\ {\rm K}$,
with and without dust or external UV fields.
The calculation is based on a one-zone model 
equipped with detailed thermal and chemical processes.
In particular, two kinds of shocks have been examined in some detail: 
(i) cold-accretion shock ($n_{\rm H,1}=4 \times 10^3\ {\rm cm^{-3}}$) and 
(ii) supernova shock ($n_{\rm H,1}=4\ {\rm cm^{-3}}$). 
Based on the obtained temperature evolution, we have estimated the fragmentation mass scales 
and discussed the condition for subsolar-mass star formation.
Our major findings can be summarized as follows:

\begin{itemize}

\item
Shock-compressed clouds contract isobarically via atomic or molecular line cooling,
until self-gravitating clumps are produced by fragmentation.
In the subsequent collapse phase, the temperature of a clump is increased due to compressional heating as well as H$_2$ formation heating,
while the density increases by a few orders of magnitude.

\item
In models without dust, no further fragmentation of a clump is expected during the collapse phase.
The clump mass is higher than $3\ {\rm M}_\odot$ in all cases~(Fig. \ref{fig:condition_Tr30}a, c).

\item
The presence of dust hardly changes the thermal evolution and the clump mass at fragmentation in the isobaric phase.
In the collapse phase, dust cooling enables these clumps to fragment again into sub-solar mass dense cores, 
as long as the metallicity is higher than $\sim 10^{-5}\ {\rm Z}_\odot$~(Fig. \ref{fig:condition_Tr30}b, d).

\item
UV irradiation affects the thermal evolution and clump mass considerably for the models with a low initial density, as in SN shocks.
In some models with enough metals~($Z \gtrsim 10^{-3}\ {\rm Z}_\odot$), the clump experiences another episode of fragmentation by metal-line cooling during the collapse phase.
This, however, only produces fragments as massive as $30\mbox{-}10^3\ {\rm M}_{\sun}$~(asterisks in Fig. \ref{fig:frag_masses}d).

\item
Even if a cloud starts contraction with different initial densities or UV strengths, 
thermal tracks converge into a common one for each metallicity
when the density is high enough for dust cooling to become effective, inducing re-fragmentation of a clump into sub-solar mass dense cores.
Therefore, the mass scale of a dense core is controlled by the amount of dust.

\end{itemize}

The evolution and fragmentation of shock-compressed clouds have also been studied by \cite{Safranek-Shrader2010},
for the case of a CA flow without considering dust grains, using a 
semi-analytical model similar to ours.
They find that when the metallicity becomes higher than $\simeq 10^{-4}\mbox{-}10^{-3}\ {\rm Z}_\odot$,
the fragment mass shows a sharp drop from $\simeq 10^5\ {\rm M}_\odot$ to $< 100\ {\rm M}_\odot$,
which agrees with our results~(the red lines in Fig. \ref{fig:frag_masses}a, c).
They also find that $\lesssim 3\ {\rm M}_\odot$ fragments form when $Z \geq 10^{-2.5}\ {\rm Z}_\odot$, while we have shown that the fragment mass never becomes lower than $\simeq 10\ {\rm M}_\odot$ in the absence of dust.
This difference comes from the fact that they do not follow the evolution after  fragmentation.
In the absence of dust, the fragment contracts while raising its temperature so rapidly that the instantaneous Jeans mass exceeds the fragment mass.
In this case, the fragment contracts in a quasi-static manner by accreting surrounding materials and keeping its mass equal to the instantaneous Jeans mass.
This boosts the fragment mass to the maximum value of the instantaneous Jeans mass during evolution, which is $\simeq 10\ {\rm M}_\odot$ when $Z \geq 10^{-2.5}\ {\rm Z}_\odot$.

Our results are based on a simple one-zone model, and any multi-dimensional effect like rotation and turbulence may modify our conclusions.
\cite{Inoue2015} studied the thermal evolution of shock-compressed metal-poor clouds using three-dimensional~(3D) hydrodynamical simulations.
They find that the shock-compressed layer becomes highly turbulent and inhomogeneous both in density and temperature, 
owing to the growth of thermal instabilities.
Their study was, however, unable to follow the evolution up to the point
where gravitationally unstable fragments form.
3D hydrodynamical calculations including self-gravity are needed to study the multi-dimensional effects on the fragment mass scales.

Magnetic fields may have played an important role in star formation, also in early galaxies.
\cite{Inoue2008, Inoue2009, Inoue2012} studied, in the context of present-day star-formation, 
the effect of magnetic fields on the evolution of a shock-compressed layer, both by  
2D and 3D magneto-hydrodynamical~(MHD) simulations.
The result depends on the angle between the magnetic field and the flow direction in the following way.
If the field line is perpendicular to the flow direction, the magnetic pressure 
balances with the ram pressure after weak compression.
The shocked gas is then hardly compressed and cools isochorically instead, lowering its pressure.
The fragment mass, calculated by the effective Jeans mass, is boosted by lower fragmentation densities 
as well as by the contribution of magnetic and turbulent pressures.
On the other hand, if the magnetic field and the flow direction are parallel to each other, the magnetic pressure 
does not work against compression, so that it will have little effect on the thermal evolution and fragmentation.
In order to see if such magnetic effects are operating in low-metallicity clouds, similar multi-dimensional MHD calculations 
are needed.

Extremely metal-poor~(EMP) stars observed in the Galactic halo provide valuable information
about the birthplace of first low-mass stars.
\cite{Schneider2012b} analyzed the origin of an EMP star SDSS J1029+1729 with a total metallicity
of $\sim 4.5 \times 10^{-5}\ {\rm Z}_\odot$, discovered by \cite{Caffau2011}.
They concluded that this star was formed in a cloud containing both metals and dust grains released by
a Pop III SN of $20\mbox{-}40\ {\rm M}_\odot$.
Although they considered a parent cloud contracting without shock-compression,
we find that this star can also be formed in a shock-compressed cloud.
More recently, \cite{Aguado2018} reported the discovery of a mega metal-poor star SDSS J0023+0303
with the iron abundance of [Fe/H] < -6.6.
Only the upper limit is obtained for the carbon abundance~([C/H] < -2.1). 
If there is indeed no carbon enhancement, the total metallicity is extremely low~($< 10^{-6.6}\ {\rm Z}_\odot$) and
dust-driven fragmentation considered here does not operate.
In such a case, we have to invoke another pathway for low-mass star formation.
Once a protostar is formed, it gains mass through the accretion from the circumstellar disc.
A multiple protostar system might be formed by disc fragmentation and some of the satellite protostars would be ejected from the system through multi-body gravitational interaction.
If the growth of a protostar is stunted before its mass exceeds $\sim {\rm M}_\odot$, a low-mass star can form even in the zero metal case~\citep{Clark_Glover2011,Greif2012,Susa2014,Chiaki2016}.
This provides a possible pathway for the formation of the star SDSS J0023+0303.

\section*{Acknowledgments}
We thank Drs Hide Yajima, Kazu Sugimura, and Gen Chiaki for fruitful discussions.
Numerical calculations are performed by the computer cluster, {\tt Draco}, supported by the Frontier Research Institute for Interdisciplinary Sciences in Tohoku University.
This work is supported in
part by MEXT/JSPS KAKENHI grants (DN:16J02951, KO:25287040, 17H01102, 17H02869).
The research leading to these results has received funding from the European Research Council under the European Union's Seventh Framework Programme (FP/2007-2013)/ERC Grant Agreement n. 306476.



\bibliographystyle{mnras}
\bibliography{ref} 








\bsp	
\label{lastpage}
\end{document}